\documentclass[twocolumn,preprintnumbers,amsmath,amssymb,superscriptaddress,longbibliography]{revtex4-2}
\usepackage{amsmath}
\usepackage{float}
\usepackage{amsthm}
\usepackage{comment}
\usepackage{amssymb}
\usepackage{subfiles}
\usepackage{bm}
\usepackage[colorlinks, linkcolor=blue, citecolor=blue, urlcolor=blue, breaklinks=true]{hyperref}
\usepackage[normalem]{ulem}
\usepackage[T1]{fontenc}
\usepackage{graphicx}
\usepackage{dcolumn}
\usepackage{amssymb,amsmath,color}
\usepackage{eqnarray}
\usepackage{booktabs}
\usepackage{mathtools}
\usepackage{amsthm}
\usepackage{dsfont}
\usepackage{cases}
\usepackage{physics}
\usepackage{subfiles}

\usepackage{subcaption}
\usepackage{caption}
\usepackage{ragged2e}
\usepackage{braket}
\usepackage[dvipsnames]{xcolor}
\usepackage{ulem} 
\usepackage[normalem]{ulem}
\usepackage[utf8]{inputenc}
\usepackage{amsmath}
\newcommand{\commentold}[1]{}
\DeclareMathSymbol{:}{\mathpunct}{operators}{"3A}
\usepackage{orcidlink}
\usepackage{amsthm}
\theoremstyle{definition}
\newtheorem{definition}{Definition}[] 

\def\be{\begin{equation}}
\def\ee{\end{equation}}
\def\bea{\begin{eqnarray}}
\def\eea{\end{eqnarray}}

\def\be{\begin{equation}}
\def\ee{\end{equation}}
\def\bal{\begin{align*}}
\def\eal{\end{align*}}
\def\bea{\begin{eqnarray}}
\def\eea{\end{eqnarray}}

\begin{document}
\date{\today}
\title{Charge-Preserving Operations in Quantum Batteries}
\author{André H. A. Malavazi\orcidlink{0000-0002-0280-0621}}
\email{andrehamalavazi@gmail.com}
\address{International Centre for Theory of Quantum Technologies, University of Gdańsk, Jana Bażyńskiego 8, 80-309 Gdańsk, Poland}
\author{Borhan Ahmadi\orcidlink{0000-0002-2787-9321}}
\email{borhan.ahmadi@ug.edu.pl}
\address{International Centre for Theory of Quantum Technologies, University of Gdańsk, Jana Bażyńskiego 8, 80-309 Gdańsk, Poland}
\author{Paweł Horodecki}
\address{International Centre for Theory of Quantum Technologies, University of Gdańsk, Jana Bażyńskiego 8, 80-309 Gdańsk, Poland}
\author{Pedro R. Dieguez\orcidlink{0000-0002-8286-2645}}
\email{dieguez.pr@gmail.com}
\address{International Centre for Theory of Quantum Technologies, University of Gdańsk, Jana Bażyńskiego 8, 80-309 Gdańsk, Poland}
\begin{abstract}
Ergotropy provides a fundamental measure of the extractable work from a quantum system and, consequently, of the maximal useful energy, or charge, stored within it. Understanding how this quantity can be manipulated and transformed efficiently is crucial for advancing quantum energy management technologies.
Here, we introduce and formalize the concepts of isoergotropic states and ergotropy-preserving operations, which reorganize the internal structure of ergotropy while keeping its total value unchanged. These ideas are illustrated for both discrete (two-level systems) and continuous-variable systems (single-mode Gaussian states). In each case, we show how ergotropy-preserving operations redistribute the respective coherent–incoherent and displacement–squeezing components.
We further examine the thermodynamic exchanges accompanying ergotropy-preserving operations, including variations in energy and entropy, and demonstrate that these transformations can be dynamically implemented through standard beam-splitter-type interactions with an auxiliary system. Finally, we discuss the practical implications of isoergotropic states and operations in optimizing charging protocols and mitigating charge loss in open quantum batteries.
\end{abstract}

\maketitle
\section{Introduction}

Quantum thermal devices designed to perform specific tasks hold significant promise for emerging energy-based technologies. The ability to precisely design and control the energetics across quantum systems has direct applications in a wide range of small-scale quantum devices. Thus, the development of machines capable of harnessing energy fluxes, performing work, and manipulating heat transport, such as quantum heat engines, refrigerators, thermal transistors, rectifiers and etc.~\cite{rossnagel2016single,peterson2019experimental,klatzow2019experimental,nie2022experimental,Manzano_2019,PhysRevE.109.064146,PhysRevB.107.075440,10.1063/5.0237842,dieguez2023thermal}, has been established as one of the main motivations behind current efforts in quantum thermodynamics.

Among such devices, quantum batteries (QBs) have emerged as a promising and powerful paradigm for energy storage. In contrast to its classical counterparts, QBs are capable of allocating a portion of their energy within their quantum degrees of freedom~\cite{PhysRevE.87.042123,camposeo2025quantum,RevModPhys.96.031001,ahmadi2025harnessing}. 
During the last decade, it has motivated a growing interest in various important aspects of QBs, including the characterization of QB architectures in a wide range of promising physical platforms~\cite{PhysRevA.109.012204,batteries8050043, PhysRevA.107.023725, PhysRevA.109.062432,Zheng_2022,PhysRevLett.131.260401,PhysRevA.110.032205,quach2022superabsorption,PhysRevA.106.042601,Cruz_2022,6bm4-8ckl}, in distinct physical contexts~\cite{PhysRevLett.132.210402,PhysRevApplied.23.024010,barra2022quantum,Shaghaghi2022,Shaghaghi2023,PhysRevA.111.042216}, and in investigating their intrinsic quantum and collective advantages~\cite{Andolina2018,PhysRevLett.124.130601,binder2015quantacell,PhysRevE.99.052106,PhysRevE.102.042111,PhysRevLett.120.117702,PhysRevLett.118.150601,Dou2021,PhysRevA.106.032212,PhysRevB.105.115405,PhysRevLett.128.140501,deMoraes_2024,10.3389/fphy.2022.1097564}.
From a practical perspective, considerable attention has been focused on developing and optimizing charging protocols~\cite{PhysRevA.107.032218,Rodríguez_2024,Hu_2022,https://doi.org/10.1002/qute.202500422,beder2025work}, and providing strategies to ensure stable charge retention in the presence of detrimental environmental effects~\cite{PhysRevA.100.043833, PhysRevE.103.042118,PhysRevA.109.052206,PhysRevE.100.032107,Mitchison2021chargingquantum,Liu2019,selfdis, stable1,Carrega_2020, PhysRevLett.122.210601,PhysRevB.99.035421,PhysRevLett.132.090401,PhysRevA.102.060201,PhysRevA.103.033715,PhysRevResearch.2.013095,bv4w-jr6q}.

In this context, the charge of a QB is commonly identified by its ergotropy, the operational figure-of-merit for quantifying the maximal useful extractable energy in closed quantum systems and thus its work capacity~\cite{A.E.Allahverdyan_2004}. 
This particular thermodynamic quantity is nonadditive and fundamentally distinct from the internal energy, but closely related to the role that free energy plays~\cite{Lobejko2021,Biswas2022extractionof}, since it implicitly incorporates the notion that energy extraction requires an external cyclic Hamiltonian control.
It is worth mentioning that different notions and extensions of ergotropy have been investigated in interacting subsystems~\cite{PhysRevA.111.012212} and open quantum systems~\cite{PhysRevA.107.012405,di2024local,PhysRevLett.133.150402}.

More recently, ergotropy has been shown to be differently distributed within the multiple internal degrees of freedom of QBs. Notably, it can be generally decomposed into two distinct contributions: an incoherent part, associated with population inversion in the energy eigenbasis, and a coherent part, arising from quantum coherence~\cite{PhysRevLett.125.180603,guha2022activation,niu2024experimental}. 
While the former can be readily identified as a purely classical contribution, the latter is genuinely quantum, i.e., without any classical analogous.
In the context of QBs, coherence has been shown to enhance energy storage~\cite{PhysRevLett.125.180603} and help mitigate ergotropy losses in open quantum batteries~\cite{bv4w-jr6q}. More recently, the coherent ergotropy of a single spin has been experimentally accessible in a nitrogen-vacancy (NV) center platform~\cite{PhysRevLett.133.180401}, elucidating its practical relevance for the current state-of-the-art technologies.
More importantly, the internal configuration of charge is not unique, and one can distribute it across all the accessible variables.
Indeed, the ergotropy stored in arbitrary Gaussian states, for instance, was shown to be divided between independent displacement and squeezing contributions~\cite{medina2024anomalous}.

Different degrees of freedom can store useful energy, so distinct states may carry the same ergotropy (charge) while differing in how that charge is internally encoded. These encodings respond unevenly to control fields and environmental noise, making states with equal charge operationally non-equivalent. This raises a natural question: which encoding is best for a given setting or thermodynamic task?
Isoergotropic (ergotropy-preserving) reorganization addresses this question. By redistributing charge among internal degrees of freedom without changing its total amount, one can refine charging protocols, enhance work extraction, propose novel charge stabilization techniques, and prepare batteries that are more robust to specific noise mechanisms—all of which are crucial for realistic, open-system implementations.
From a thermodynamic perspective, such operations also enable new cycles whose strokes reshuffle charge internally at a fixed budget. 
Understanding—and exploiting—the interplay between these internal compositions of ergotropy is therefore both fundamental and practically important for the design of energy-based quantum devices.

In this work, we formalize the set $\mathcal{L}$ of ergotropically equivalent quantum states and the class $\mathcal{K}$ of operations that preserve total ergotropy while redistributing its internal components. Our procedure covers both CPTP channels and non-selective measurements.
We then analyze the fundamental trade-offs among internal ergotropy components for two representative battery models: discrete two-level systems (TLSs) and continuous variable (CV) single-mode Gaussian states.
On the implementation side, we show that these ergotropy-preserving operations can be realized dynamically with standard beam-splitter interactions to an auxiliary system, available in current platforms.
Finally, we illustrate how isoergotropic states and operations aid practice: they enable the refinement of charging protocols and mitigate charge loss by exploiting an ergotropic counterpart of the Mpemba effect.

The remainder of this article is structured as follows. In Sec.~\ref{DefErgo} we recall the definition of ergotropy and introduce isoergotropic states and ergotropy-preserving operations. In Secs. \ref{TLS} and \ref{GaussianStates}, we apply these ideas to quantum batteries built from discrete and continuous-variable cells, modeled by two-level systems and single-mode Gaussian states, respectively. Each section is self-contained: we identify the corresponding isoergotropic states, specify the ergotropy-preserving operations, and describe how to reorganize ergotropy dynamically within the internal degrees of freedom using a standard beam-splitter coupling.
In Sec.~\ref{Application}, we illustrate how the formalism developed here applies to relevant scenarios for quantum batteries, including charging and discharging protocols.
Finally, Sec. \ref{DiscussionSection} summarizes the main conclusions and outlines directions for future work.
\begin{figure}
    \centering
    \includegraphics[width=1\columnwidth]{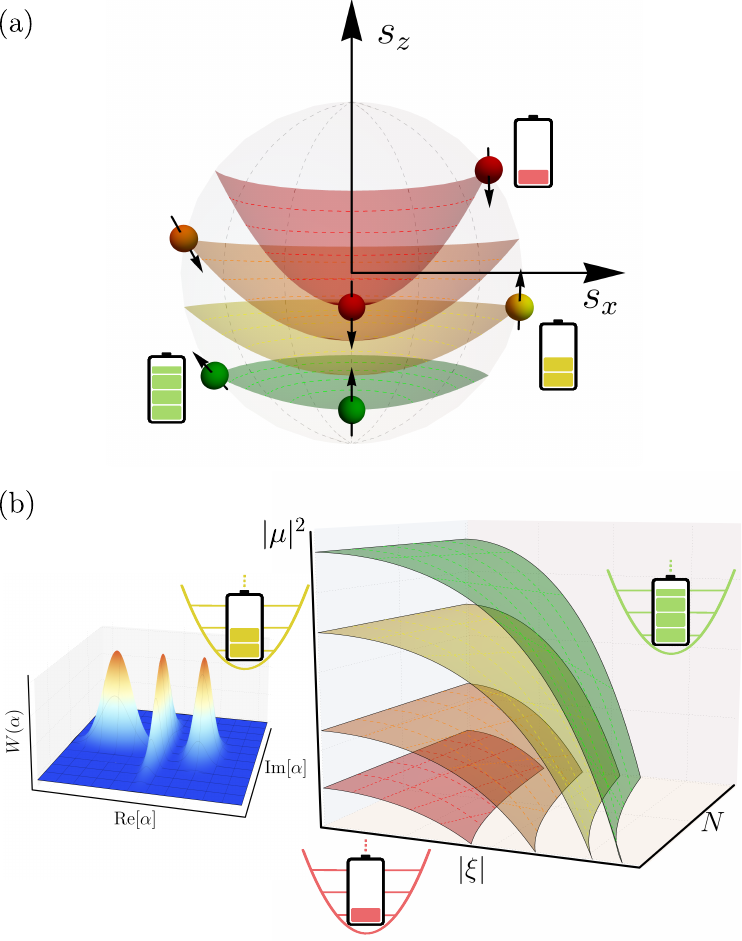}
    \caption{Isoergotropic states for (a) a single two-level quantum battery and (b) a single-mode Gaussian oscillator. Each colored surface at the (a) Bloch sphere (b) parameter space, corresponds to states with fixed charge but distinct internal configurations of ergotropy $\mathcal{R}$, with red and green being respectively low and high charge. The Wigner functions $W(\alpha)$ represent Gaussian states with the same charge and thermal occupation $N$ but different displacement $\mu$ and squeezing $\xi$ components of $\mathcal{R}$ \cite{Serafini}. \justifying}
    \label{Diagram}
\end{figure}
%
%
%
%
%
\begin{figure*}
    \center
    \includegraphics[width=2\columnwidth]{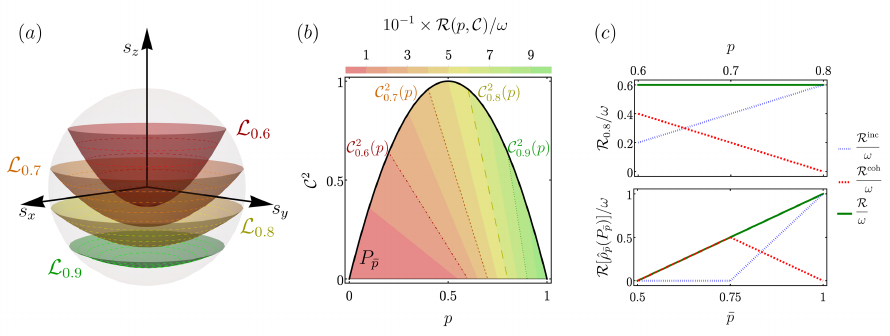}
    \caption{(a) Isoergotropic surfaces at the Bloch sphere for distinct values of $\bar{p}$. (b) Ergotropy profile in terms of $p$ and $\mathcal{C}^{2}\in[0,4p(1-p)]$. The isoergotropic states within $\mathcal{L}_{\bar{p}}$ are represented by the colored (a) surfaces within the Bloch sphere (b) lines $\mathcal{C}_{\bar{p}}^{2}(p)=8(\bar{p}-p)P_{\bar{p}}$. The black curve in (b) characterize the isoergotropic pure states $\hat{\rho}_{\bar{p}}(P_{\bar{p}},\theta)=|\Psi_{\bar{p},\theta}\rangle\langle\Psi_{\bar{p},\theta}|$ (surface of the Bloch sphere) relative to the incoherent states $\hat{\varrho}_{\bar{p}}=\left(1-p\right)|g\rangle\langle g|+p|e\rangle\langle e|$ (at $s_z$ axis). (c) Top: Ergotropy distribution between the internal (in)coherent components along the path described by $\mathcal{C}_{0.8}^{2}(p)$. Bottom: Ergotropy distribution along the pure states $\hat{\rho}_{\bar{p}}(P_{\bar{p}},\theta)$ represented by the black curve. \justifying}
    \label{fig2}
\end{figure*}
\section{Ergotropy and Isoergotropic operations}\label{DefErgo}
Let us assume a $d$-dimensional quantum system described by a density matrix $\hat{\rho} = \sum_{j}^{d} p_{j} |j\rangle \langle j|\in\mathcal{L}$ and Hamiltonian $\hat{H} = \sum_{k}^{d} E_{k} |E_{k}\rangle \langle E_{k}|$. 
As a quantum battery, its charge is commonly characterized by the maximum extractable energy through a unitary cyclic Hamiltonian control. This quantity, also known as ergotropy, can be written as~\cite{A.E.Allahverdyan_2004}
\begin{equation}\label{ErgotropyDef}
    \mathcal{R}[\hat{\rho}] \coloneqq \textrm{Tr} \left[ \left( \hat{\rho} - \hat{\rho}_{p} \right) \hat{H} \right] = \sum_{k,j}^{d} E_{k} p_{j} \left( |\langle E_{k} | j \rangle|^{2} - \delta_{k,j} \right),
\end{equation}
where $\hat{\rho}_{p} = \sum_{i}^{d} p_{i} |E_{i}\rangle \langle E_{i}|$ is the passive state relative to $\hat{\rho}$, and the canonical spectral ordering $p_{1} \geq p_{2} \geq \dots \geq p_{d}$ and $E_{1} \leq E_{2} \leq \dots \leq E_{d}$ were assumed. Note that Eq.~\eqref{ErgotropyDef} can equivalently be obtained by identifying the unitary $\hat{V}$ that reorders the populations ${p_{j}}$ so as to minimize the system’s energy,
$\hat{\rho} \mapsto \hat{\rho}_{p} = \hat{V}\hat{\rho}\hat{V}^{\dagger}$.
In general, the total ergotropy of a system may be distributed among distinct internal degrees of freedom. Consequently, different states can possess identical ergotropy—hence the same stored charge—while exhibiting distinct internal configurations.
Formally, we introduce the following definition:
\begin{definition}[]
    Given an active state $\hat{\varrho}$ with ergotropy $\mathcal{\bar{R}}=\mathcal{R}[\hat{\varrho}]$, the set of \textit{isoergotropic (iso-$\mathcal{R}$) states} relative to $\hat{\varrho}$ is defined as $\bar{\mathcal{L}}\coloneqq\left\{ \hat{\rho}\in\mathcal{L}\,|\,\mathcal{R}[\hat{\rho}]=\mathcal{\bar{R}}\right\} $.
\end{definition}
Figure~\ref{Diagram} illustrates such states for (a) a two-level system and (b) a single-mode Gaussian state \cite{Serafini}. In panel (a), the colored surfaces on the Bloch sphere, and in panel (b), the colored sheets in the $\big(|\mu|^{2},|\xi|,N\big)$ space, collect states with the same total ergotropy but different internal compositions of charge. 
The inset Wigner functions $W(\alpha)$ illustrate distinct Gaussian states with equal total charge and thermal occupation $N$ but different displacement $\mu$ and squeezing $\xi$.
Transformations connecting states within a given isoergotropic surface correspond to {\it ergotropy-preserving} operations, which redistribute the internal composition of stored work while keeping its total value invariant.
Unitaries that commute with the Hamiltonian, $[\hat{U}, \hat{H}] = 0$ (e.g., free evolution), constitute trivial instances of such operations, leaving the internal configuration unchanged.
We therefore define:
\begin{definition}
    Given a state $\hat{\varrho}\in\bar{\mathcal{L}}$, iso-$\mathcal{R}$ operations are CPTP (measurements) maps satisfying $\mathcal{K}:\bar{\mathcal{L}}\mapsto\bar{\mathcal{L}}$.
\end{definition}

These operations provide the ``free transformations" in a prospective resource theory of ergotropy at a fixed budget.
The resource corresponds to the configuration of stored work within a given level set $\bar{\mathcal{L}}$,
while the iso-$\mathcal{R}$ maps $\mathcal{K}$ represent the admissible reshufflings that preserve the total ergotropy.
Below, we instantiate this for discrete (TLS) and continuous-variable Gaussian batteries.
%
%
%
%
%
\section{Two-level systems}\label{TLS}
We begin with a TLS described by the Hamiltonian $\hat{H} = \omega|e\rangle \langle e|$ and by the density operator $\hat{\rho}  =\left(1-p\right)|g\rangle\langle g|+p|e\rangle\langle e|+\mathcal{C}\left(e^{i\frac{\theta}{2}}|g\rangle\langle e|+h.c.\right)/2$, where $h.c.$ denotes the Hermitian conjugate, $\theta \in[0,4\pi]$ and $\mathcal{C}\coloneqq\sum_{k\neq j}|\langle k|\hat{\rho}|j\rangle|$, with $\mathcal{C}^{2}\in[0,4p(1-p)]$, is the $l_{1}$ norm of coherence in the energy eigenbasis~\cite{RevModPhys.89.041003}. 

In general, Eq.~\eqref{ErgotropyDef} for ergotropy can be split into a contribution due to population inversion and a contribution due to coherences (both defined in the Hamiltonian eigenbasis). Define the incoherent state
$\hat\rho_{\mathrm{diag}}=\sum_{k}^{d}\langle E_k|\hat\rho|E_k\rangle\,|E_k\rangle\!\langle E_k|$ and set
$\mathcal R^{\mathrm{inc}}[\hat\rho]\coloneqq\mathcal R[\hat\rho_{\mathrm{diag}}]$, $\mathcal R^{\mathrm{coh}}[\hat\rho]\coloneqq\mathcal R[\hat\rho]-\mathcal R^{\mathrm{inc}}[\hat\rho]$. Then the ergotropy is fully parameterized by $(p,\mathcal C)$: $\mathcal R(p,\mathcal C)=\mathcal R^{\mathrm{inc}}(p)+\mathcal R^{\mathrm{coh}}(p,\mathcal C)$,
with~\cite{PhysRevLett.125.180603,guha2022activation,niu2024experimental,bv4w-jr6q}
\begin{equation}\label{IncoherentErgoFormula}
    \mathcal R^{\mathrm{inc}}(p)=\omega(2p-1)\,\big[1-\Theta(1/2-p)\big],
\end{equation}
and
\begin{equation}\label{CoherentErgoFormula}
    \mathcal R^{\mathrm{coh}}(p,\mathcal C)=\frac{\omega}{2}\!\left(\psi-\sqrt{\psi^{2}-\mathcal C^{2}}\right),
\end{equation}
where $\Theta(x)$ is the Heaviside step function ($\Theta(x)=1$ for $x\ge0$,
$\Theta(x)=0$ otherwise), $\mu[\hat\rho]\coloneqq\mathrm{Tr}[\hat\rho^{2}]$ is the purity, and $\psi\coloneqq\sqrt{2\mu[\hat\rho]-1}$.
Both components are independent of the phase $\theta$; variations of $\theta$ thus generate a trivial family of isoergotropic states. Moreover,
$\mathcal R^{\mathrm{coh}}(p,\mathcal C)=0$ for incoherent states
($\mathcal C=0$), while $\mathcal R^{\mathrm{inc}}(p)$ is nonzero iff $p>1/2$ (the active regime).
\subsection{Isoergotropic states}
Given an active incoherent state $\hat{\varrho}_{\bar{p}}=\left(1-\bar{p}\right)|g\rangle\langle g|+\bar{p}|e\rangle\langle e|$ with $\bar{p}\in]1/2,1]$ with ergotropy $\mathcal{R}_{\bar{p}}\coloneqq\mathcal{R}(\bar{p},0)=\mathcal{R}^{\textrm{inc}}(\bar{p})$, the corresponding isoergotropic set is
\begin{equation}
    \mathcal{L}_{\bar{p}}\coloneqq\left\{ \hat{\rho}_{\bar{p}}\in\mathcal{L}\,|\,\mathcal{R}[\hat{\rho}_{\bar{p}}(p,\theta)]=\mathcal{R}_{\bar{p}}\right\},
\end{equation}
A convenient parametrization of the states in $\mathcal{L}_{\bar p}$ is
\begin{equation}
    \hat{\rho}_{\bar{p}}(p,\theta)=\left(\begin{array}{cc}
1-p & \mathcal{C}_{\bar{p}}(p)e^{i\frac{\theta}{2}}/2\\
\mathcal{C}_{\bar{p}}(p)e^{-i\frac{\theta}{2}}/2 & p
\end{array}\right),
\end{equation}
where $\mathcal{C}_{\bar{p}}^{2}(p)=8(\bar{p}-p)P_{\bar{p}}$ and $P_{\bar{p}}\leq p\leq\bar{p}$, with $P_{\bar{p}}=2\bar{p}-1$ and $\theta\in[0,4\pi]$.
As extreme states, one has the mixed reference state $\hat{\varrho}_{\bar{p}}$ (by construction) for $p=\bar{p}$ and its relative isoergotropic pure state $\hat{\rho}_{\bar{p}}(P_{\bar{p}},\theta)=|\Psi_{\bar{p},\theta}\rangle\langle\Psi_{\bar{p},\theta}|$ for $p=P_{\bar{p}}$, with $|\Psi_{\bar{p},\theta}\rangle=\sqrt{2(1-\bar{p})}|g\rangle+e^{-i\theta/2}\sqrt{2\bar{p}-1}|e\rangle$.
Thus, for each fixed $\bar{p}$, the total ergotropy is redistributed between the incoherent and coherent parts, such that 
\begin{equation}
    \mathcal{R}_{\bar{p}}=\mathcal{R}^{\textrm{inc}}(p)+\mathcal{R}^{\textrm{coh}}(p,\mathcal{C}_{\bar{p}}),
\end{equation}
while remaining constant.
Figure~\ref{fig2} (a) shows the Bloch sphere representation, with $s_{k}=\textrm{Tr}[\hat{\sigma}_{k}\hat{\rho}_{\bar{p}}(p)]$ with $k=x,y,z$, and the isoergotropic surfaces generated by different values of $\bar{p}$. For each $\bar{p}$, the incoherent states $\hat{\varrho}_{\bar{p}}$ lies on the $s_z$ axis, while the isoergotropic pure state $\hat{\rho}_{\bar{p}}(P_{\bar{p}},\theta)$ lies on the surface of the sphere.
Figure~\ref{fig2} (b) displays the ergotropy profile on the $p-\mathcal{C}^{2}$ surface; the colored curves correspond to the same isoergotropic sets as in (a), and the black curve at $p=P_{\bar p}$ marks the pure states.
Finally, Fig.~\ref{fig2} (c) depicts how the internal ergotropy splits into coherent and incoherent components along two representative paths: (Top) $\mathcal{C}_{0.8}^{2}(p)$ and (Bottom) $P_{\bar{p}}$.
\subsubsection*{Thermodynamics}

\begin{figure}
    \centering
    \includegraphics[width=1\columnwidth]{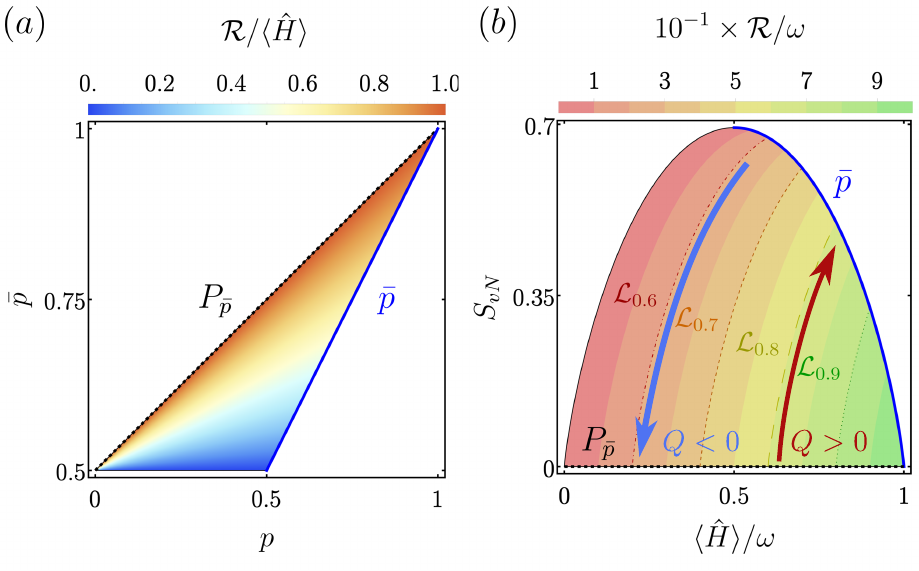}
    \caption{(a) Ratio $\mathcal{R}/\langle H \rangle = (2\bar{p} - 1)/p$ on the $(p, \bar{p})$-plane. For each $\bar{p}$, the ratio decreases monotonically with $p$ from $1$ at the pure point $p = P_{\bar{p}}$ to $(2\bar{p} - 1)/\bar{p}$ at the incoherent point $p = \bar{p}$. (b) Energy-entropy diagram on $\mathcal{L}_{\bar{p}}$: $\langle \hat{H} \rangle = \omega p$ and $S_{vN}(p) = \mathcal H_2(2\bar{p} - p)$ increase monotonically with $p$. Arrows indicate heat $Q = \omega(p' - p)$ absorbed (right, red) or released (left, blue) when moving along $\mathcal{L}_{\bar{p}}$. Black dashed: $p = P_{\bar{p}}$ (pure, $S_{vN} = 0$); blue solid: $p = \bar{p}$ (incoherent, $S_{vN} = \mathcal H_2(\bar{p})$).\justifying}
    \label{FigQubitThermo}
\end{figure}

Having identified the isoergotropic set $\mathcal{L}_{\bar{p}}$, we now examine how key thermodynamic quantities vary along it.
In this sense, note that, despite having the same ergotropy, different iso-$\mathcal{R}$ states necessarily contain different internal energies, unambiguously identified as~\cite{e24111645}
\begin{equation}
    \langle\hat{H}\rangle=\textrm{Tr}[\hat{H}\hat{\rho}_{\bar{p}}(p,\theta)]=\omega p.
\end{equation}
So, any iso-$\mathcal{R}$ transformation $\rho_{\bar{p}}(p, \theta) \mapsto \rho_{\bar{p}}(p', \theta')$ induces the following energetic change
\begin{equation}
\Delta U \equiv \Delta \langle H \rangle = \omega(p' - p).
\end{equation}
If the Hamiltonian is fixed (no explicit time dependence), no work is performed $W = \text{Tr}[\rho \, \Delta H] = 0$~\cite{Alicki_1979,ahmadi2023work}, and the first law of thermodynamics $\Delta U = Q + W$ reduces to
\begin{equation}
Q = \omega(p' - p).
\end{equation}
Therefore, moving between the iso-$\mathcal{R}$ states requires the supply or release of heat. Notice that, for any $\bar{p}$, states closer to the center of the Bloch sphere are more energetic, i.e., $p$ closer to $\bar{p}$ and the z-axis. Given the invariance of ergotropy, this also implies the proportion of energy stored as charge decreases, that is, $\mathcal{R}/\langle\hat{H}\rangle<1$ for $p<P_{\bar{p}}$. To illustrate this, Fig.~\ref{FigQubitThermo} (a) shows the ratio between charge and internal energy for all values of $p$ and $\bar{p}$.
Additionally, Fig.~\ref{FigQubitThermo} (b) depicts the internal energy vs. entropy diagram for all isoergotropic states. The von Neumann entropy is written as 
\begin{equation}
   S_{vN}[\hat{\rho}_{\bar{p}}(p,\theta)]= \mathcal H_2(2\bar{p}-p),
\end{equation}
with $\mathcal H_2(x) = -x\ln{x} - (1-x)\ln{(1-x)}$ being the binary entropy function, varies across the isoergotropic surfaces. 
The blue lines highlight the mixed states along the z-axis with $p=\bar{p}$, while the black dashed lines show the region with pure states $p=P_{\bar{p}}$ at the surface of the Bloch sphere, with $S_{vN}[\hat{\rho}_{\bar{p}}(P_{\bar{p}},\theta)]=0$.
Consequently, there are no nontrivial operations that simultaneously conserve ergotropy and are isentropic or energy-preserving: along $\mathcal L_{\bar p}$, for fixed $\hat{H}$, nontrivial iso-$\mathcal R$ transformations necessarily involve heat exchange and entropy change; they are not implementable by system-only unitaries, but arise naturally as thermal operations or as reduced dynamics of an energy-conserving global unitary.
%
%
%
%
%
\subsection{Ergotropy preserving operations}
\begin{figure}
    \centering
    \includegraphics[width=0.6\columnwidth]{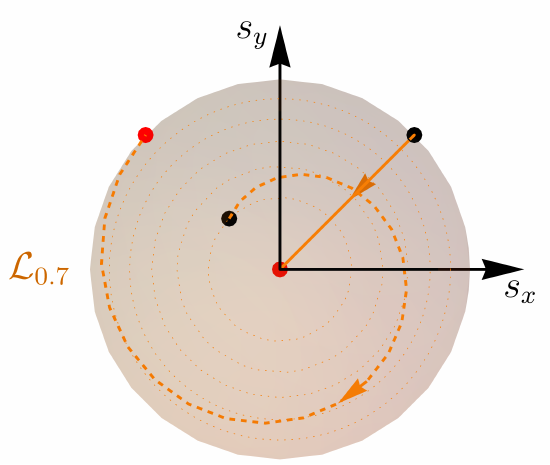}
    \caption{Isoergotropic operations. The continuous and dashed lines indicate two distinct ergotropy-preserving trajectories along $\mathcal{L}_{0.7}$ from the black to the red points.\justifying}
    \label{Fig3}
\end{figure}

It is clear that the tradeoff between coherent and incoherent components of ergotropy within different states of $\mathcal{L}_{\bar p}$ implies that iso-$\mathcal R$ operations are, in general, neither incoherent nor purity-preserving (and, as noted above, not entropy-preserving either). In fact, incoherent operations preserve only the incoherent part of ergotropy. Given a state $\hat\rho_{\bar p}(p,\theta)\in\mathcal{L}_{\bar p}$, we call iso-$\mathcal R$ operations the CPTP maps (or measurement instruments) that satisfy
\begin{equation}
\mathcal{K}:\ \mathcal{L}_{\bar p}\longrightarrow\mathcal{L}_{\bar p},
\qquad
\hat\rho_{\bar p}(p',\theta')=\mathcal{K}\!\left[\hat\rho_{\bar p}(p,\theta)\right],
\end{equation}
with $p'\in[P_{\bar p},\bar p]$.
Let $\hat\rho_{\bar p}(p',\theta')=(1-p_e)\,|\psi_g\rangle\!\langle\psi_g|+p_e\,|\psi_e\rangle\!\langle\psi_e|$ be the spectral decomposition of the target state, where $p_e=2\bar p-p'$ and $\{|\psi_g\rangle,|\psi_e\rangle\}$ are its eigenvectors. A convenient Kraus representation $\{\hat{\mathcal K}_j\}$ of a channel $\mathcal{K}_{\bar p,p',\theta'}$ that connects $\hat\rho_{\bar p}(p,\theta)\to\hat\rho_{\bar p}(p',\theta')$ is
\begin{equation}\label{Kraus}
\begin{aligned}
\hat{\mathcal K}_1&=\sqrt{p_e}\,|\psi_e\rangle\!\langle\psi_e|, &
\hat{\mathcal K}_2&=\sqrt{p_e}\,|\psi_e\rangle\!\langle\psi_g|,\\
\hat{\mathcal K}_3&=\sqrt{1-p_e}\,|\psi_g\rangle\!\langle\psi_g|, &
\hat{\mathcal K}_4&=\sqrt{1-p_e}\,|\psi_g\rangle\!\langle\psi_e|,
\end{aligned}
\end{equation}
such that
\begin{equation}
\hat\rho_{\bar p}(p',\theta')=\sum_{j=1}^{4}\hat{\mathcal K}_j\,\hat\rho_{\bar p}(p,\theta)\,\hat{\mathcal K}_j^\dagger.
\end{equation}
Equation~\eqref{Kraus} has the structure of the generalized amplitude damping channel (GADC)~\cite{PhysRevA.102.012401}. Equivalently, the same map can be realized as a qubit thermal channel (QTC, or thermal attenuator)~\cite{Rosati2018},
\begin{equation}
\mathcal{E}_{\bar p,p',\theta'}[\hat\rho]=\mathrm{Tr}_{\mathrm{Aux}}\!\left[\hat U\left(\hat\rho\otimes\hat\tau_{\bar p}\right)\hat U^\dagger\right],
\end{equation}
with $\hat\tau_{\bar p}\equiv\hat\rho_{\bar p}(p',\theta')$ being an auxiliary state and
\begin{equation}\label{unitary}
\hat U=
\begin{pmatrix}
1 & 0 & 0 & 0\\
0 & 0 & 1 & 0\\
0 & 1 & 0 & 0\\
0 & 0 & 0 & 1
\end{pmatrix},
\end{equation}
a SWAP gate.
To illustrate iso-$\mathcal R$ operations, Fig.~\ref{Fig3} shows two distinct ergotropy-preserving trajectories along $\mathcal{L}_{0.7}$ generated from Eq.~\eqref{Kraus}. The black (red) dots mark the initial (final) states. By choosing the parameters appropriately, any path on the Bloch sphere within $\mathcal{L}_{0.7}$ can be implemented.

\subsubsection{Iso-$\mathcal{R}$ realization via POVM}

Iso-$\mathcal{R}$ operations can also be realized through selective measurements (POVM elements).
A single selective measurement is sufficient to realize an isoergotropic operation that maps an incoherent two-level system active state to any other state within the same isoergotropic family.
Indeed, consider the incoherent active state
\begin{equation}
\hat{\varrho}_{\bar{p}}=(1-\bar{p})\ket{g}\bra{g}+\bar{p}\ket{e}\bra{e},
\end{equation}
and its corresponding isoergotropic family $\mathcal{L}_{\bar{p}}$.
When the desired target is a \emph{pure} member of $\mathcal{L}_{\bar{p}}$, a rank-one selective measurement suffices. Let the single measurement operator be
\begin{equation}
\hat{M}=\ket{\phi_{\mathrm{out}}}\bra{\phi_{\mathrm{in}}},
\end{equation}
which, acting on $\hat{\varrho}_{\bar{p}}$, yields the (unnormalized) post-measurement state
\begin{equation}
\hat{M}\hat{\varrho}_{\bar{p}}\hat{M}^\dagger
=\bra{\phi_{\mathrm{in}}}\hat{\varrho}_{\bar{p}}\ket{\phi_{\mathrm{in}}}
\ket{\phi_{\mathrm{out}}}\bra{\phi_{\mathrm{out}}}.
\end{equation}
After renormalization, the output is pure, such that
\begin{equation}
\hat{\rho}'=\ket{\phi_{\mathrm{out}}}\bra{\phi_{\mathrm{out}}}.
\end{equation}
To remain within $\mathcal{L}_{\bar{p}}$, the output state $\ket{\phi_{\mathrm{out}}}$ must be the normalized eigenvector of the target isoergotropic element,
\begin{equation}
\ket{\phi_{\mathrm{out}}}=\ket{\Psi_{\bar{p}}(p,\theta)}=\frac{\left(\frac{\mathcal{C}_{\bar{p}}(p)}{2p}e^{i\theta/2}\ket{g}+\ket{e}\right)}{\sqrt{1+\dfrac{\mathcal{C}_{\bar{p}}^{2}(p)}{4p^{2}}}},
\end{equation}
so that $\hat{\rho}'=\hat{\rho}_{\bar{p}}(p,\theta)$ is a pure isoergotropic state.

The success probability of this transformation is
\begin{equation}
\mathcal{P}_{\mathrm{succ}}(p)=
\bra{\phi_{\mathrm{in}}}\hat{\varrho}_{\bar{p}}\ket{\phi_{\mathrm{in}}},
\end{equation}
which is maximized by choosing $\ket{\phi_{\mathrm{in}}}$ as the eigenvector of $\hat{\varrho}_{\bar{p}}$ associated with its largest eigenvalue, i.e. $\ket{e}$.
Hence, the optimal Kraus operator reads
\begin{equation}
\hat{M}_{\mathrm{opt}}(p,\theta)=
\ket{\Psi_{\bar{p}}(p,\theta)}\bra{e},
\end{equation}
and the maximal success probability follows as
\begin{equation}
\mathcal{P}_{\mathrm{succ}}^{\max}(p)=\bar{p}.
\end{equation}

For a general \emph{mixed} target in $\mathcal{L}_{\bar{p}}$, a rank-one operator is no longer sufficient, as it always yields a pure output.
Nevertheless, a single selective measurement with a higher-rank Kraus operator can achieve the transformation in one step.
Specifically, consider the ansatz
\begin{equation}
\hat{M}=\sqrt{q}\sqrt{\hat{\rho}_{\bar{p}}(p',\theta')}
\hat{\varrho}_{\bar{p}}^{-1/2},
\end{equation}
with $q\in[0,1]$ chosen such that $\hat{M}^\dagger\hat{M}\leq\hat{\mathbb{I}}$. Indeed, defining
\begin{equation}
\hat{A}\coloneqq\hat{\varrho}_{\bar{p}}^{-1/2}\,\hat{\rho}_{\bar{p}}(p',\theta')\,\hat{\varrho}_{\bar{p}}^{-1/2},
\end{equation}
which is positive. The necessary and sufficient condition for $\hat M^\dagger\hat M\le\hat{\mathbb I}$ is
\begin{equation}
q\,\lambda_{\max}(\hat{A})\le 1.
\end{equation}
By construction,
\begin{equation}
\hat{M}\hat{\varrho}_{\bar{p}}\hat{M}^\dagger
=q\,\hat{\rho}_{\bar{p}}(p',\theta'),
\end{equation}
and the normalized post-measurement state (upon the success outcome) is precisely the mixed isoergotropic element $\hat{\rho}_{\bar{p}}(p',\theta')$.

For the present incoherent input,
one may compute $A$ explicitly:
\begin{equation}
\hat{A}=\begin{pmatrix}
\frac{1-p'}{1-\bar p} & \frac{\mathcal C_{\bar p}(p')}{2\sqrt{(1-\bar p)\bar p}}\,e^{i\theta'/2}\\[6pt]
\frac{\mathcal C_{\bar p}(p')}{2\sqrt{(1-\bar p)\bar p}}\,e^{-i\theta'/2} & \frac{p'}{\bar p}
\end{pmatrix}.
\end{equation}
An explicit form for the achievable success probability is
\begin{equation}\label{eq:q_explicit}
q_{\max}=\frac{2}{\,T+\sqrt{T^2-4D}\,},
\end{equation}
where
\begin{equation}
\begin{aligned}
T &= \frac{1-p'}{1-\bar p}+\frac{p'}{\bar p},\\[4pt]
D &= \frac{(1-p')p' - \big(\tfrac{\mathcal C_{\bar p}(p')}{2}\big)^2}{(1-\bar p)\bar p}.
\end{aligned}
\end{equation}

Finally, if one is allowed to reset the input to $\hat{\varrho}_{\bar p}$ after any failed attempt, the trials become independent and identical with per-trial success probability $q_{\max}$. In that operational setting the cumulative success probability after $N$ attempts is the geometric (Bernoulli) form 
\begin{equation}
\mathcal P_{\mathrm{succ}}^{(N)}=1-(1-q_{\max})^{N},
\end{equation}
which tends to unity as $N\to\infty$.
Thus, a single selective measurement suffices to generate \emph{any} state—pure or mixed—within the isoergotropic family $\mathcal{L}_{\bar{p}}$ starting from the incoherent active state $\hat{\varrho}_{\bar{p}}$.

%
%
%
%
%
%
%
%
%
%
%
%
%
%
%
%
%
\begin{figure}
    \center
    \includegraphics[width=1\columnwidth]{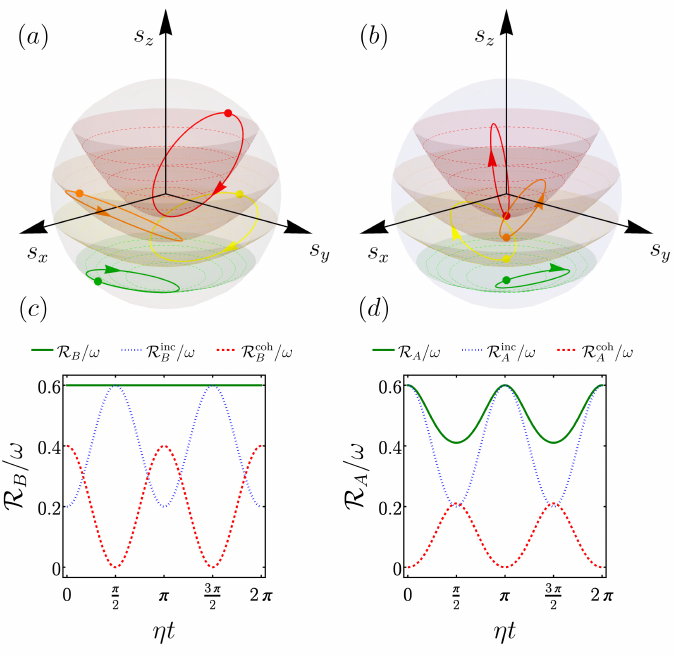}
    \caption{Top: Bloch sphere representation of the (a) battery and (b) auxiliary TLS dynamics induced by $\hat{\mathcal{U}}(t)$, assuming $\hat{\rho}_{B}(0)=\hat{\rho}_{\bar{p}}(P_{\bar{p}},\theta)$ and $\hat{\rho}_A(0)=\hat{\rho}_{\bar{p}}(\bar{p},\phi)$ (colored dots) as initial states, and $\omega=\eta=1$. The isoergotropic curves $\mathcal{L}_{\bar{p}}$ for different values of $\bar{p}$ are represented. Bottom: Ergotropy and internal components dynamics for the (c) battery and (d) auxiliary system for $\bar{p}=0.8$. \justifying}
    \label{fig8}
\end{figure}
\subsection{Dynamic ergotropy conversion}
We now show how iso-$\mathcal R$ operations can be engineered autonomously by
coupling a quantum battery ($B$) to an auxiliary system ($A$).
Consider two TLS with total Hamiltonian
$\hat H=\hat H_B+\hat H_A+\hat V_{AB}$, where
$\hat H_{A,B}=\omega_{A,B}\,|e\rangle\langle e|_{A,B}$ and
\begin{equation}\label{int1}
\hat V_{AB}=i\eta\big(\hat\sigma_+^B\!\otimes\!\hat\sigma_-^A
-\hat\sigma_-^B\!\otimes\!\hat\sigma_+^A\big),
\end{equation}
with coupling strength $\eta$. For simplicity we take resonant TLSs,
$\omega_A=\omega_B=\omega$. If the initial local states are
$\hat\rho_B(0)$ and $\hat\rho_A(0)$, then
$\hat\rho_{BA}(t)=\hat{\mathcal U}(t)\big(\hat\rho_B(0)\!\otimes\!\hat\rho_A(0)\big)\hat{\mathcal U}^\dagger(t)$ and
$\hat\rho_{B(A)}(t)=\mathrm{Tr}_{A(B)}[\hat\rho_{BA}(t)]$, with
\begin{equation}\label{TLSUnitary}
\hat{\mathcal U}(t)=e^{-it\omega}
\begin{pmatrix}
e^{it\omega} & 0 & 0 & 0\\
0 & \cos(\eta t) & \sin(\eta t) & 0\\
0 & -\sin(\eta t) & \cos(\eta t) & 0\\
0 & 0 & 0 & e^{-it\omega}
\end{pmatrix}.
\end{equation}
Notice that Eq.~\eqref{TLSUnitary} has the same swap block structure as Eq.~\eqref{unitary}.
Prepare both systems on the same isoergotropic curve $\mathcal L_{\bar p}$ with $\hat\rho_B(0) = \hat\rho_{\bar p}(P_{\bar p},\theta)$ and
$\hat\rho_A(0) = \hat\rho_{\bar p}(\bar p,\phi)$. Then the battery state at time $t$ reads
\begin{equation}
\hat\rho_B(t)\equiv \hat\rho_{\bar p}\big(p_B(t),\,\theta+2\omega t\big)\in\mathcal L_{\bar p},
\end{equation}
where
\begin{equation}
p_B(t)=\tfrac12\big[(\bar p-1)\cos(2\eta t)+\bar p+P_{\bar p}\big]\in[P_{\bar p},\bar p].
\end{equation}
Figure~\ref{fig8}(a,b) shows the corresponding trajectories on the Bloch sphere for the battery and the auxiliary (for several $\bar p$ and $\omega=\eta=1$). The battery remains on the same isoergotropic surface, whereas the auxiliary generally does not. Panels (c,d) display the dynamics of total ergotropy and its coherent/incoherent components for both subsystems: the auxiliary ergotropy oscillates with period $\eta T=\pi$, while the battery cyclically converts coherent ergotropy into incoherent ergotropy and back.

Although the total ergotropy of the composite is conserved because $[\hat{\mathcal U}(t),\hat H]=0$, the local decrease of auxiliary ergotropy quantifies the internal energetic cost of the conversion on the battery. In realistic settings, the interaction strength $\eta$ can be modulated in time to tailor this conversion. If only local ergotropy is available, the full energetic cost comprises both the auxiliary’s ergotropy consumption and the control required to modulate the interaction. If global access to the composite ergotropy is available, only the external control cost remains relevant.

\begin{figure}
    \center
    \includegraphics[width=\columnwidth]{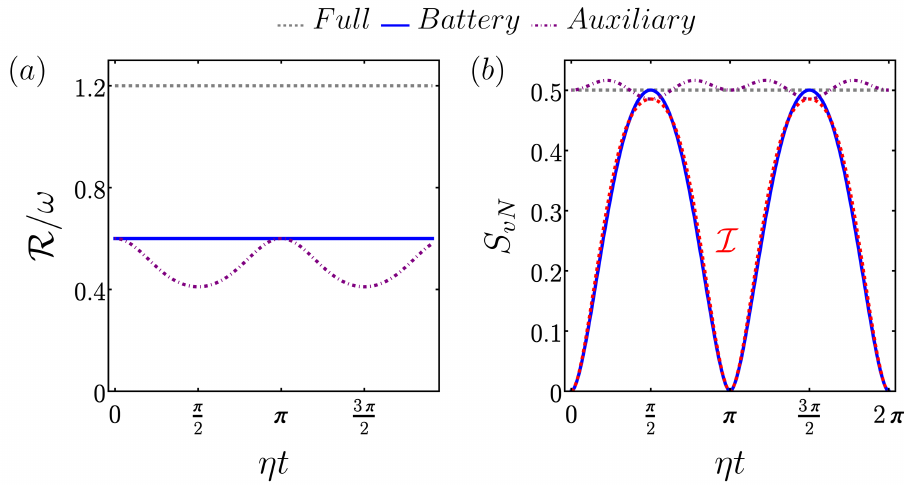}
    \caption{(a) Charge and (b) von Neumann entropy dynamics induced by $\hat{\mathcal{U}}(t)$ for $\bar{{p}}=0.8$. The entropy and ergotropies of the full system (computed without the interaction, i.e., $\hat{H}_A+\hat{H}_B$) and the battery are invariant during the time-evolution. The ergotropy of the auxiliary system oscillation is accompanied by the conversion of coherent ergotropy to incoherent. Such oscillation is also observed by the local entropies $S_{vN}$ and mutual information $\mathcal{I}$.\justifying}
    \label{fig9}
\end{figure}

Figure~\ref{fig9} further reports (a) ergotropy and (b) the von Neumann entropies $S_{vN}$ and the mutual information $\mathcal I = S_{vN}[\hat\rho_B]+S_{vN}[\hat\rho_A]-S_{vN}[\hat\rho_{BA}]$ during the interaction for $\bar p=0.8$. The auxiliary’s ergotropy oscillations are in resonance with changes in local entropies. The growth of mutual information signals the build-up of correlations, accompanying a decrease (increase) of the battery’s coherent ergotropy (the auxiliary’s ergotropy).
\subsection{Extension}\label{Extension}
Naturally, one can extend the iso-$\mathcal R$ analysis to multi–cell batteries composed of $n$ TLSs and identify suitable ergotropy–preserving operations implemented by local or global transformations. As a simple and relevant two–cell example, consider the family of $X$-states with Hamiltonian $\hat H=\omega\,|e\rangle\langle e|_{1}\otimes\hat{\mathds 1}_{2}
+\hat{\mathds 1}_{1}\otimes\omega\,|e\rangle\langle e|_{2}$ and a single scalar parameter
$q\in[0,1]$~\cite{behzadi2018thermodynamic,gomes2022realism,bv4w-jr6q}:
\begin{equation}\label{StateX}
    \hat{\rho}_{X}(q)=\left(\begin{array}{cccc}
q/2 & 0 & 0 & q^{2}-q/2\\
0 & q(1-q) & 0 & 0\\
0 & 0 & (1-q)^{2} & 0\\
q^{2}-q/2 & 0 & 0 & q/2
\end{array}\right).
\end{equation}
\begin{figure}
    \centering
    \includegraphics[width=0.85\columnwidth]{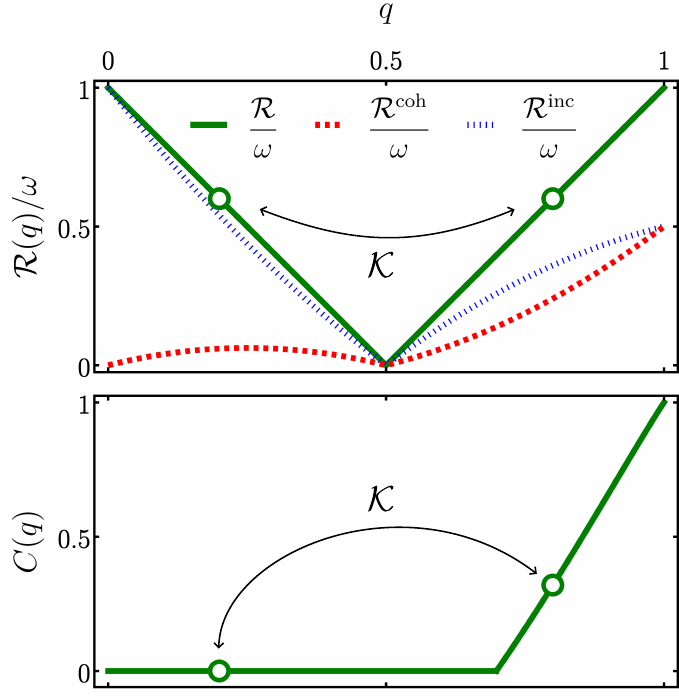}
    \caption{Top: Ergotropy profile as a function of $q$. The total ergotropy satisfies $\mathcal{R}(q)=\mathcal{R}(1-q)$ for all $q$; the dotted curves show the coherent and incoherent components. Bottom: Concurrence $C(q)$. Every uncorrelated state with $0\le q\le 0.3$ has a correlated isoergotropic partner with $0.7\le q\le 1$. The arrow illustrates an isoergotropic operation $\mathcal{K}$ mapping $q\mapsto 1-q$.\justifying}
    \label{Fig4}
\end{figure}
Equation~\eqref{StateX} can be obtained as the output of an optimal unitary applied to two Gibbs states that maximizes quantum correlations~\cite{gomes2022realism}—including entanglement and discord~\cite{behzadi2018thermodynamic}—and local uncertainty. For this $X$-family, the concurrence is
\begin{equation}
C(q)=\max\!\Big\{0,\ 2q^{2}-q-2(1-q)\sqrt{q(1-q)}\Big\}.
\end{equation}
While entanglement vanishes at a finite critical temperature~\cite{behzadi2018thermodynamic,nielsen2000quantum,PhysRevA.59.1799}, other quantum signatures decay gradually with temperature
\cite{behzadi2018thermodynamic,gomes2022realism}. 
\begin{figure}
    \center
    \includegraphics[width=0.85\columnwidth]{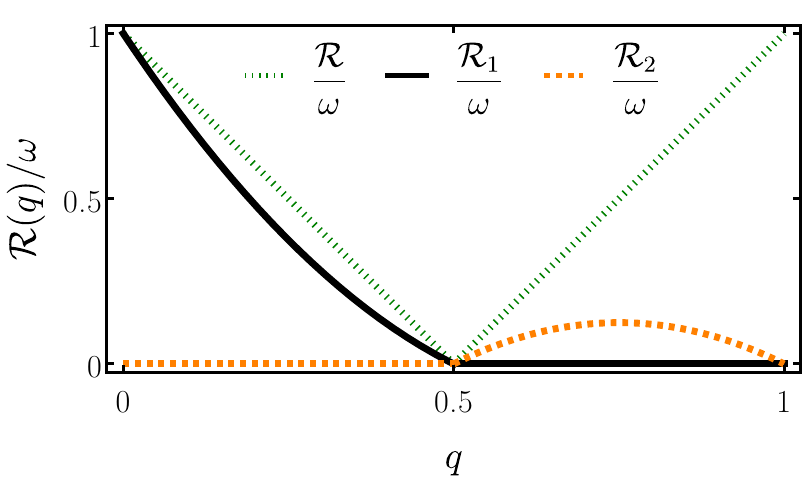}
    \caption{Ergotropy profile as a function of $q$ for cell $1$ (black) and $2$ (orange). The total ergotropy is shown in green.\justifying}
    \label{Ext2}
\end{figure}
The ergotropy within the bipartition is
\begin{equation}
\mathcal R(q)=\omega\,|1-2q|,
\qquad
\mathcal R^{\mathrm{inc}}(q)=\omega\,|(q-2)(q-\tfrac12)|,
\end{equation}
hence $\mathcal R(q)=\mathcal R(1-q)$ for all $q$.
Thus, $\hat\rho_X(q)$ and $\hat\rho_X(1-q)$ are isoergotropic, even though their entanglement may differ significantly~\cite{bv4w-jr6q}. 
Figure~\ref{Fig4} (top) highlights the symmetric ergotropy profile and the split between coherent and incoherent contributions; Fig.~\ref{Fig4} (bottom) shows the concurrence, including entanglement sudden death. In particular, every uncorrelated state with $0\le q\le 0.3$ has a correlated
isoergotropic partner with $0.7\le q\le 1$. A notable pair is the pure product $\hat\rho_X(0)$ and the maximally entangled
$\hat\rho_X(1)$: both have the same ergotropy, with
$\mathcal R^{\mathrm{inc}}(1)=\mathcal R^{\mathrm{coh}}(1)=\omega/2$.   
Moreover, one finds that the internal energy is constant for all $q$ while the von Neumann entropy is symmetric about $q=\tfrac12$ (with a maximum at $q=\tfrac12$): $\langle\hat{H}\rangle=\textrm{Tr}[\hat{H}\hat{\rho}_{X}(q)]=\omega$ and $S_{vN}(q)=S_{vN}(1-q)$. Consequently, the fraction of energy stored as charge decreases near $q=1/2$ and for the isoergotropic transformation $q\mapsto 1-q$ one has $\Delta U=\Delta S_{vN}=0$ and $W=-Q$.

An explicit iso-$\mathcal R$ operation connecting the partners can be written as
\begin{equation}
   \mathcal{K}\left[\hat{\rho}_{X}(q)\right]=\hat{V}\textrm{Tr}_{\textrm{Aux}}\left[\hat{U}\left(\hat{\rho}_{X}(q)\otimes\hat{\tau}_{\textrm{Aux}}\right)\hat{U}^{\dagger}\right]\hat{V}^{\dagger},
\end{equation}
where $\hat U$ is the $4\times 4$ SWAP, the auxiliary state is
$\hat\tau_{\mathrm{Aux}}=\hat\tau_1\otimes\hat\tau_2$ with
$\hat\tau_k=(1-q)|g\rangle\langle g|_{k}+q|e\rangle\langle e|_{k}$, and $\hat V=\hat v_2\hat v_1$ with
\begin{equation}\label{}
\begin{aligned}
\hat{v}_{1}&=\ket{gg}\bra{gg}+\ket{ge}\bra{ge}+\ket{ee}\bra{eg}+\ket{eg}\bra{ee},\\\hat{v}_{2}&=\ket{\phi_{+}}\bra{gg}+\ket{ge}\bra{ge}+\ket{eg}\bra{eg}+\ket{\phi_{-}}\bra{ee},
\end{aligned}
\end{equation}
and $\ket{\phi_{\pm}}=(\ket{gg}\pm\ket{ee})/\sqrt{2}$. This construction realizes the isoergotropic mapping $q\mapsto 1-q$ by a thermal auxiliary system plus a global unitary and a fixed postprocessing unitary $\hat V$. Note that the mapping $q\mapsto 1-q$ cannot be implemented deterministically by local operations and classical communications (LOCC) \cite{PhysRevLett.83.436} for $q\in(0,1)$: for $q<\tfrac12$ it generally increases entanglement (e.g., the concurrence of $\hat\rho_X(q)$ is smaller than that of $\hat\rho_X(1-q)$), which is forbidden under LOCC; accordingly, our construction employs a thermal ancilla and a global unitary.

In this sense, it is natural to consider isoergotropic transformations according to the LOCC paradigm for work extraction in correlated systems similar to the framework proposed in Ref.~\cite{PhysRevLett.89.180402}, where local heat engines can act on each subsystem and classical communication is allowed. In our setting, one may similarly ask whether a transformation that is non–isoergotropic when constrained locally can nevertheless become globally isoergotropic—or at least less non–isoergotropic—once classical coordination and suitable ancillas are considered. This motivates defining an LOCC–iso-$\mathcal R$ class of operations that preserve the total ergotropy of the composite while acting locally, and introducing an “ergotropy deficit” to quantify the difference between globally preserved ergotropy and what is retained by purely local means. A systematic characterization of the minimal nonlocal resources (shared randomness, classical vs. quantum communication, ancilla temperature) required to uplift locally non–isoergotropic processes to globally isoergotropic ones is an open direction for our future work.

From a thermodynamic perspective, the coupling to the auxiliary implements a heat exchange $Q=\omega(2q-1)$, while the subsequent unitary $\hat{V}$ performs coherent work $W=-Q$, so that the net energy change vanishes: $\Delta U = Q+W=0$. Moreover, the initial and final von Neumann entropies coincide, $S_{vN}\!\left[\hat\rho_X(q)\right]=S_{vN}\!\left[\hat\rho_X(1-q)\right]$, so the transformation $q\mapsto 1-q$ is entropy-preserving overall. It is worth mentioning that in the SWAP step, the heat exchanged between the composite system and the auxiliary does not go to the correlations inside the composite system since we have no interaction within them: $\langle H\rangle = \langle H_1\rangle+\langle H_2\rangle$. This heat is accounted for entirely by changes in the local energies.

As for the local ergotropies, the bipartite state in Eq.~\eqref{StateX}, the reduced states
$\hat\rho_k(q)=\mathrm{Tr}_{\bar k}[\hat\rho_X(q)]$ ($k=1,2$; $\bar k=2,1$) are diagonal with $p_1(q)=1-\tfrac{3}{2}q+q^{2}$ and $p_2(q)=\tfrac{3}{2}q-q^{2}$ being the excited-state populations of the subsystems 1, 2 respectively. Figure~\ref{Ext2} plots $\mathcal R_1$ (black), $\mathcal R_2$ (orange), and $\mathcal R$ (green):
for $q<\tfrac{1}{2}$ only cell 2 carries charge; for $q>\tfrac{1}{2}$ only cell 1 does. The total ergotropy $\mathcal R(q)=\omega|1-2q| \ge \mathcal R_1(q)+\mathcal R_2(q)$, the excess being due to correlations/coherences. Under the isoergotropic mapping $q\mapsto 1-q$, each cell exchanges heat equal to its energy change (no local work), giving
\begin{equation}
Q_{1}=\omega\left(q-\dfrac{1}{2}\right),\quad
Q_2=-Q_1.
\end{equation}
Hence, for $q>\tfrac{1}{2}$ cell 1 absorbs heat ($Q_1>0$) while cell 2 releases heat ($Q_2<0$); the roles reverse for $q<\tfrac{1}{2}$. The global energy is constant and the total ergotropy is preserved.

%
%
\begin{figure*}
    \center
    \includegraphics[width=2\columnwidth]{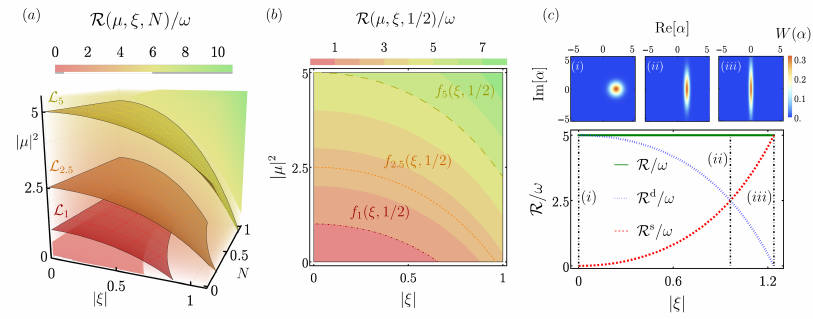}
    \caption{(a) Ergotropy profile. The isoergotropic surfaces $\mathcal{L}_{|\bar{\mu}|^2}$ are shown for different values of $\bar{\mu}$. (b) Ergotropy profile in terms of $|\mu|^2-|\xi|$ for $N=1/2$. The colored lines highlight the iso-$\mathcal{R}$ curves $f_{\bar{\mu}}(\xi,1/2)$ for the same values shown in (a).
    (c) Bottom: Ergotropy distribution along the isoergotropic curve $f_{5}(\xi,1/2)$. The black vertical lines at (i) $|\xi|=0$, (ii) $|\xi|\approx0.96$ and (iii) $|\xi|\approx1.24$ highlight the points where (i) $\mathcal{R}^{\textrm{s}}=0$, (ii) $\mathcal{R}^{\textrm{d}}=\mathcal{R}^{\textrm{s}}$ and (iii) $\mathcal{R}^{\textrm{d}}=0$, respectively. Top: Wigner function $W(\alpha)$ for each of the cases mentioned, assuming $\theta=0$ and $\phi=\pi$.\justifying}
    \label{figGaussian}
\end{figure*}
%
\section{Bosonic Gaussian states}\label{GaussianStates}
Consider a single-mode bosonic field of frequency $\omega$ with Hamiltonian
$\hat H=\omega(\hat a^\dagger \hat a+\tfrac12)$, where $\hat a$ and $\hat a^\dagger$ satisfy $[\hat a,\hat a^\dagger]=1$.
Assume the system is in a Gaussian state of the form
\begin{equation}\label{GaussianState}
\hat\rho=\hat D(\mu)\,\hat S(\xi)\,\hat\pi(N)\,\hat S^\dagger(\xi)\,\hat D^\dagger(\mu),
\end{equation}
where $\hat\pi(N)$ is a thermal state with occupation $N=(e^{\beta\omega}-1)^{-1}$, $\hat D(\mu)=\exp(\mu\hat a^\dagger-\mu^*\hat a)$ is the displacement operator, and $\hat S(\xi) = \exp\!\big[\tfrac12(\xi\hat a^{\dagger 2}-\xi^*\hat a^2)\big]$ is the squeezing operator, with complex parameters $\mu=|\mu|e^{i\theta}$ and $\xi=|\xi|e^{i\phi}$. Any single-mode Gaussian state admits the decomposition in Eq.~\eqref{GaussianState}~\cite{Adam01061995}.
Equivalently, Gaussian states are fully characterized by first and second moments: the mean vector $\boldsymbol d=(\langle \hat a\rangle,\langle \hat a^\dagger\rangle)^{T}=(\mu,\mu^*)^{T}$ and the covariance matrix $\boldsymbol\Theta=(N+\tfrac12)\,\boldsymbol F(\xi,\phi)$, with
\begin{equation}
\boldsymbol F(\xi,\phi)=
\begin{pmatrix}
\cosh(2|\xi|) & e^{i\phi}\sinh(2|\xi|)\\
e^{-i\phi}\sinh(2|\xi|) & \cosh(2|\xi|)
\end{pmatrix}.
\end{equation}
A convenient phase-space representation is given by the Wigner function, defined as
\begin{equation}
W(\alpha)=\frac{1}{\pi\sqrt{|\boldsymbol\Theta|}}\,
\exp\!\left[-\tfrac12\,(\boldsymbol\alpha-\boldsymbol d)^\dagger
\boldsymbol\Theta^{-1}(\boldsymbol\alpha-\boldsymbol d)\right],
\end{equation}
with $\boldsymbol\alpha=(\alpha,\alpha^*)$ and $|\boldsymbol\Theta|=\det\boldsymbol\Theta$.
Both displacement and squeezing contribute independently to the total ergotropy, so the stored charge can be parameterized by $(\mu,\xi,N)$ as
$\mathcal R(\mu,\xi,N)=\mathcal R^{\mathrm d}(\mu)+\mathcal R^{\mathrm s}(\xi,N)$, with~\cite{PhysRevLett.125.180603,medina2024anomalous}
\begin{equation}\label{ErgoGaussian}
\begin{aligned}
\mathcal{R}^{\mathrm{d}}(\mu)&=\omega\,\frac{|\boldsymbol{d}|^{2}}{2}=\omega|\mu|^{2},\\\mathcal{R}^{\mathrm{s}}(\xi,N)&=\omega\!\left(\frac{\mathrm{Tr}\,\boldsymbol{\Theta}}{2}-\sqrt{|\boldsymbol{\Theta}|}\right)=\omega\Big(N+\tfrac{1}{2}\Big)\\&\quad\times\big[\cosh(2|\xi|)-1\big].
\end{aligned}
\end{equation}
Note that $N$ enters only the squeezing contribution. Moreover, Eq.~\eqref{ErgoGaussian} is independent of the phases $\theta$ and $\phi$, which provides a trivial class of iso-$\mathcal R$ states.

\subsection{Isoergotropic states}
Given an active Gaussian state of the form $\hat{\varrho}_{\bar{\mu}}=\hat{D}(\bar{\mu})\hat{\pi}(0)\hat{D}^{\dagger}(\bar{\mu})$, with ergotropy $\mathcal{R}_{\bar{\mu}}\coloneqq\mathcal{R}(\bar{\mu},0,0)=\mathcal{R}^{\textrm{d}}(\bar{\mu})$, the set of isoergotropic Gaussian states relative to $\hat{\varrho}_{\bar{\mu}}$ is written as
\begin{equation}
    \mathcal{L}_{|\bar{\mu}|^2}\coloneqq\left\{ \hat{\rho}_{\bar{\mu}}\in\mathcal{L}\,|\,\mathcal{R}[\hat{\rho}_{\bar{\mu}}(\mu,\xi,N)]=\mathcal{R}_{\bar{\mu}}\right\},
\end{equation}
for $0\leq|\mu|\leq|\bar{\mu}|$, $|\xi|\geq0$ and $N\geq 0$.
These states can be written as $\hat{\rho}_{\bar{\mu}}(\mu,\xi,N)=\hat{D}(\mu)\hat{S}(\xi)\hat{\pi}(N)\hat{S}^{\dagger}(\xi)\hat{D}^{\dagger}(\mu)$, where $\mu=\sqrt{f_{\bar{\mu}}(\xi,N)}e^{i\theta}$ with
\begin{equation}
    f_{\bar{\mu}}(\xi,N)=|\bar{\mu}|^{2}-\left(N+\frac{1}{2}\right)\left[\cosh\left(2|\xi|\right)-1\right],
\end{equation}
and $(\xi,N)$ satisfying $f_{\bar{\mu}}(\xi,N)\geq0$.
Figure~\ref{figGaussian} shows the ergotropy profile in terms of parameters $(\mu,\xi,N)$. In (a), is highlighted the isoergotropic surfaces $\mathcal{L}_{1}$, $\mathcal{L}_{2.5}$ and $\mathcal{L}_{5}$. Figure~\ref{figGaussian} (b) shows the surface $N=1/2$. The dashed lines indicate the surfaces shown in (a). The bottom part of Fig.~\ref{figGaussian} (c) presents the internal ergotropy distribution along the iso-ergotropic curve represented by $f_{5}(\xi,1/2)$. The black vertical lines highlight the points where (i) $\mathcal{R}^{\textrm{s}}=0$, (ii) $\mathcal{R}^{\textrm{d}}=\mathcal{R}^{\textrm{s}}$ and (iii) $\mathcal{R}^{\textrm{d}}=0$. The Wigner function $W(\alpha)$ for each of these cases is presented in the plots at the top.
\subsubsection*{Thermodynamics}

\begin{figure}
    \centering
    \includegraphics[width=1\columnwidth]{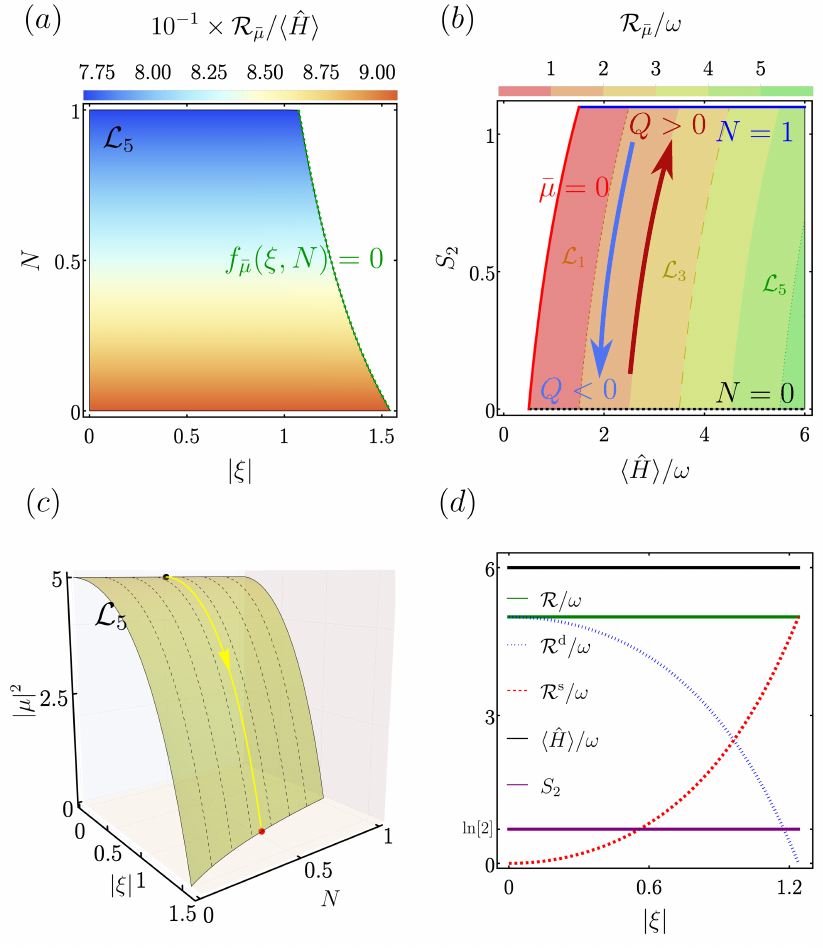}
    \caption{(a) Charge–energy ratio on $\mathcal{L}_{5}$:
    $\mathcal{R}_{\bar\mu}/\langle \hat H\rangle = \mathcal{R}_{\bar\mu}/\big(\mathcal{R}_{\bar\mu}+\omega(N+\tfrac12)\big)$. It decreases with $N$ (purely thermal energy) and is independent of $\xi$. Green dashed: $f_{\bar\mu}(\xi,N)=0$. (b) Internal energy vs. Rényi-2 entropy vs. $\mathcal{R}_{\bar\mu}$ for several $N$ and $\bar\mu$; black (blue): $N=0$ ($N=1$). Red: $\bar\mu=0$ (no ergotropy). Arrows indicate heat $Q = \omega\,\Delta N$ absorbed (right, red) or released (left, blue) when moving along $\mathcal{L}_{|\bar{\mu}|^{2}}$. (c) Fixed-$N$ lines on $\mathcal{L}_{5}$; yellow: isothermal path ($N=0.5$). (d) Along fixed-$N$ paths, $Q=\Delta S_2=0$ and
    $\langle \hat H\rangle$ and $\mathcal{R}_{\bar\mu}$ are invariant while
    $\mathcal{R}^{\mathrm s}$ and $\mathcal{R}^{\mathrm d}$ reshuffle. \justifying}
    \label{FigGaussianThermo}
\end{figure}

As in the TLS case, isoergotropic states on the same level set can differ in their
thermodynamic properties. For single–mode Gaussian states with Hamiltonian
$\hat{H}=\omega(\hat a^\dagger \hat a+\tfrac12)$, the internal energy reads
\begin{equation}\label{IntEnergy}
\langle \hat H\rangle
=\mathrm{Tr}\!\big[\hat H\,\hat\rho_{\bar\mu}(\mu,\xi,N)\big]
=\mathcal{R}^{\mathrm s}(\xi,N)+\mathcal{R}^{\mathrm d}(\mu)
+\omega\!\left(N+\tfrac12\right).
\end{equation}
Along an iso-$\mathcal R$ manifold $\mathcal{L}_{|\bar\mu|^2}$ the total ergotropy
$\,\mathcal{R}_{\bar\mu}=\mathcal{R}^{\mathrm s}+\mathcal{R}^{\mathrm d}\,$ is fixed, so
\[
\langle \hat H\rangle
=\mathcal{R}_{\bar\mu}+\omega\!\left(N+\tfrac12\right),
\]
which is independent of the squeezing parameter $\xi$. Consequently, for fixed $|\bar\mu|^2$ the ratio of charge to energy depends only on the thermal component:
\[
\frac{\mathcal{R}_{\bar\mu}}{\langle \hat H\rangle}
=\frac{\mathcal{R}_{\bar\mu}}{\mathcal{R}_{\bar\mu}+\omega\!\left(N+\tfrac12\right)}.
\]
Figure~\ref{FigGaussianThermo}(a) shows this behavior on $\mathcal{L}_{5}$ as a function of $N$ and $\xi$; the green dashed curve marks the zero set of
$f_{\bar\mu}(\xi,N)$. We quantify mixedness by the Rényi-2 entropy
$S_2(\rho)=(1-2)^{-1}\ln \mathrm{Tr}[\rho^2]$~\cite{PhysRevLett.109.190502}, for which
\begin{equation}\label{Entropy}
S_{2}\!\big[\hat{\rho}_{\bar{\mu}}(\mu,\xi,N)\big]=\ln\left(1+2N\right),
\end{equation}
i.e., $S_2$ depends only on $N$ and is invariant under displacement and squeezing.
Figure~\ref{FigGaussianThermo}(b) plots the $(\langle \hat H\rangle, \,S_2, \,\mathcal{R}_{\bar\mu})$ diagram for different $N$ and $\bar\mu$; black (blue) curves correspond to $N=0$ ($N=1$), and the red line highlights $\bar\mu=0$ (no ergotropy: all energy/entropy is thermal).
Differentiating Eq. \eqref{IntEnergy} and using $d\mathcal{R}^{\mathrm s}+d\mathcal{R}^{\mathrm d}=0$ on $\mathcal{L}_{|\bar\mu|^2}$ yields
\[
d\langle \hat H\rangle=\omega\,dN.
\]
With $\hat{H}$ fixed (no explicit time dependence) the first law gives $W=0$ and therefore
\[
Q=\Delta U=\omega\,\Delta N.
\]
Hence, reshuffling ergotropy between displacement and squeezing at fixed $N$
is thermodynamically neutral on the subsystem:
\[
dN=0\quad\Rightarrow\quad Q=0,\ \ d S_2=0,\ \ d\mathcal{R}_{\bar\mu}=0.
\]
Only changes in the thermal occupancy $N$ carry heat ($Q=\omega\,d N$) and entropy ($d S_2=2\,dN/(1+2N)$ under Eq. \eqref{Entropy}). Figure~\ref{FigGaussianThermo}(c) shows fixed-$N$ lines on $\mathcal{L}_{5}$; the yellow curve is an isothermal trajectory ($N=0.5$).
Moving from the black to the red point along this line leaves $\langle \hat H\rangle$, $S_2$, and $\mathcal{R}_{\bar\mu}$ unchanged, while the internal composition $\,\mathcal{R}^{\mathrm s}\leftrightarrow\mathcal{R}^{\mathrm d}\,$ is reconfigured, as illustrated in Fig.~\ref{FigGaussianThermo}(d).
%
%
%
\subsection{Ergotropy-preserving operations}

Restricting to Gaussian states, we likewise restrict to Gaussian iso-$\mathcal R$ operations. Given an initial state $\hat\rho_{\bar\mu}(\mu,\xi,N)$, an iso-$\mathcal R$ Gaussian map
$\mathcal K:\mathcal L_{|\bar\mu|^{2}}\to\mathcal L_{|\bar\mu|^{2}}$ produces
\begin{equation}
\hat\rho_{\bar\mu}(\mu',\xi',N')=\mathcal K\!\left[\hat\rho_{\bar\mu}(\mu,\xi,N)\right],
\end{equation}
with total ergotropy $\mathcal R_{\bar\mu}$ unchanged.
Gaussian channels are fully specified by their action on the first moments
$\boldsymbol d$ and covariance matrix $\boldsymbol\Theta$~\cite{Serafini,brask2021gaussian}.
Therefore, ergotropy invariance under $(\mu,\xi,N)\mapsto(\mu',\xi',N')$ is guaranteed by
\begin{equation}
\boldsymbol\Theta \mapsto \boldsymbol\Theta'=\Big(N'+\tfrac12\Big)\boldsymbol F(\xi',\phi'),\quad
\boldsymbol d\mapsto \boldsymbol d'=(\mu',\mu'^*)^{T},
\end{equation}
with the constraint $\mu'=\sqrt{\,f_{\bar\mu}(\xi',N')\,}\,e^{i\theta'}$, so that $\mathcal R^{\mathrm d}(\mu')+\mathcal R^{\mathrm s}(\xi',N')=\mathcal R_{\bar\mu}$. Operationally, such a map can be realized as (i) a thermalization step that sets the target occupation $N'$, followed by (ii) a squeezing $\hat{S}(\xi')$ and (iii) a displacement $\hat{D}(\mu')$. The first step may be implemented by a single-mode thermal attenuator of transmissivity $\eta$, for which
\begin{equation}
\boldsymbol d\mapsto \sqrt{\eta}\,\boldsymbol d,\qquad
\boldsymbol\Theta\mapsto \eta\,\boldsymbol\Theta+(1-\eta)\Big(N'+\tfrac12\Big)\mathds{1}_{2},
\end{equation}
while $\hat{S}(\xi')$ updates $\boldsymbol\Theta$ and $\hat{D}(\mu')$ updates $\boldsymbol d$. Equivalently—and closer to our two-level construction—one can append an auxiliary single-mode isoergotropic Gaussian state
$\hat\tau_{\mathrm{Aux}}=\hat\rho_{\bar\mu}(\mu',\xi',N')$ with moments
$(\boldsymbol d',\boldsymbol\Theta')$, realize a beamsplitter/SWAP interaction, and trace out the auxiliary mode~\cite{Rosati2018,PhysRevA.102.012401}. Writing the joint first moments $\boldsymbol D=(\boldsymbol d,\boldsymbol d')^{T}$ and joint covariance
$\boldsymbol\Xi=\boldsymbol\Theta\oplus\boldsymbol\Theta'$, the idealized SWAP block acts as
\begin{equation}
\begin{aligned}
\boldsymbol\Xi=
\begin{pmatrix}
\boldsymbol\Theta & \boldsymbol 0\\
\boldsymbol 0 & \boldsymbol\Theta'
\end{pmatrix}
\ &\longmapsto\
\hat U\,\boldsymbol\Xi\,\hat U^\dagger=
\begin{pmatrix}
\boldsymbol\Theta' & \boldsymbol 0\\
\boldsymbol 0 & \boldsymbol\Theta
\end{pmatrix},\\
\boldsymbol D=(\mu,\mu^*,\mu',\mu'^*)^{T}\ &\longmapsto\ \hat U\,\boldsymbol D=(\mu',\mu'^*,\mu,\mu^*)^{T},
\end{aligned}
\end{equation}
with
\begin{equation}
\hat U=
\begin{pmatrix}
0 & 0 & 1 & 0\\
0 & 0 & 0 & 1\\
1 & 0 & 0 & 0\\
0 & 1 & 0 & 0
\end{pmatrix},
\end{equation}
which swaps the two modes. Tracing out the auxiliary leaves the target
isoergotropic state $\hat\rho_{\bar\mu}(\mu',\xi',N')$.

\begin{figure}
    \center
    \includegraphics[width=0.8\columnwidth]{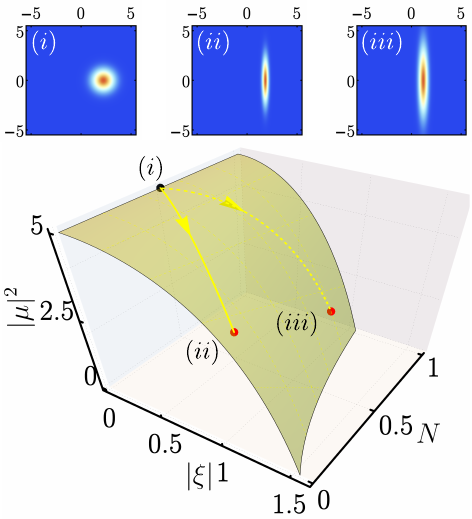}
    \caption{Isoergotropic Gaussian operations. The continuous and dashed lines indicate two distinct ergotropy-preserving trajectories along $\mathcal{L}_{5}$ from the black to the red points. It was assumed $\phi=\phi^\prime=\pi$, $\theta=\theta^\prime=0$ (i) $N=0.5$, $|\xi|=0$; (ii) $N^\prime=0.1$, $|\xi^\prime|=1$; (iii) $N^\prime=0.8$, $|\xi^\prime|=1$.\justifying}
    \label{figGaussianTraj}
\end{figure}

Figure~\ref{figGaussianTraj} illustrates two different trajectories induced by isoergotropic Gaussian operations along $\mathcal{L}_{5}$, together with the Wigner functions of the depicted states. The black dot (i) marks the initial state, while the red dots (ii)–(iii) indicate two distinct final states.
%
%
%


\subsubsection{Iso-$\mathcal{R}$ realization via selective Gaussian measurements}

Let the input be an arbitrary element of the isoergotropic family
\begin{equation}
\hat{\rho}_{\bar\mu}(\mu,\xi,N)=\hat{\mathcal U}(\mu,\xi)\,\hat\pi(N)\,\hat{\mathcal U}^\dagger(\mu,\xi),
\end{equation}
with $\hat{\mathcal U}(\mu,\xi)=\hat D(\mu)\hat S(\xi)$, $\hat\pi(N)=\sum_{n\ge 0} p_n(N)\ket{n}\bra{n}$ and
\begin{equation}
p_n(N)=\frac{N^n}{(N+1)^{n+1}}.
\end{equation}
The eigenvectors of $\hat{\rho}_{\bar\mu}(\mu,\xi,N)$ are $\hat{\mathcal U}(\mu,\xi)\ket{n}$ with eigenvalues $p_n(N)$. In particular the principal eigenvalue is
\begin{equation}
\lambda_{\max}\big(\hat\rho_{\bar\mu}(\mu,\xi,N)\big)=p_0(N)=\frac{1}{N+1}.
\end{equation}
For the isoergotropy reference given by the displaced vacuum, the desired pure target is
\begin{equation}
\hat{\rho}_{\rm tar}=\ket{\bar\mu}\bra{\bar\mu},\qquad \ket{\bar\mu}=\hat D(\bar\mu)\ket{0}.
\end{equation}
Choose the rank–one Kraus
\begin{equation}
\hat M=\ket{\bar\mu}\bra{\phi},
\end{equation}
where $\ket{\phi}$ is any normalized vector. Acting on the input yields the unnormalized post-measurement operator
\begin{equation}
\hat M\,\hat\rho_{\bar\mu}(\mu,\xi,N)\,\hat M^\dagger
= \bra{\phi}\hat\rho_{\bar\mu}(\mu,\xi,N)\ket{\phi}\ \ket{\bar\mu}\bra{\bar\mu}.
\end{equation}
After renormalization, the conditional post-measurement state is exactly the pure target:
\begin{equation}
\frac{\hat M\hat\rho_{\bar\mu}\hat M^\dagger}{\operatorname{Tr}[\hat M\hat\rho_{\bar\mu}\hat M^\dagger]}
=\ket{\bar\mu}\bra{\bar\mu}.
\end{equation}
The success probability for this outcome equals
\begin{equation}
\mathcal P_{\mathrm{succ}}(\phi)=\bra{\phi}\hat\rho_{\bar\mu}(\mu,\xi,N)\ket{\phi}.
\end{equation}
Maximizing over $\ket\phi$ yields the principal eigenvalue of the input:
\begin{equation}
\mathcal P_{\mathrm{succ}}^{\max}=\lambda_{\max}\big(\hat\rho_{\bar\mu}(\mu,\xi,N)\big)=p_0(N)=\frac{1}{N+1},
\end{equation}
and the optimal vector is
\begin{equation}
\ket{\phi_{\mathrm{opt}}}=\hat{\mathcal U}(\mu,\xi)\ket{0},
\end{equation}
so that the optimal Kraus reads
\begin{equation}
\hat M_{\rm opt}=\ket{\bar\mu}\bra{\phi_{\mathrm{opt}}}.
\end{equation} 
Analogously to the qubit case, this shows that selective measurements can be employed in Gaussian systems to transform one state into another while preserving the ergotropy.
%
%
%
%
%
%
\begin{figure}
    \center
    \includegraphics[width=1\columnwidth]{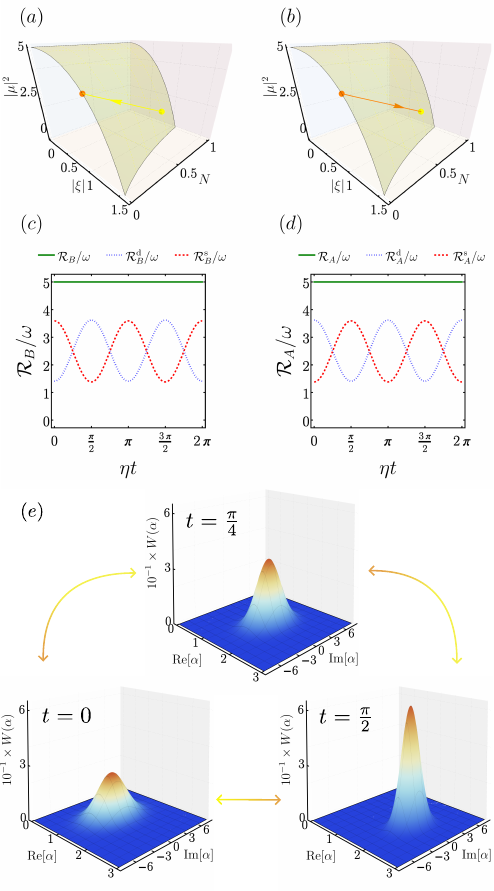}
    \caption{Top: Parameter space representation of the (a) battery and (b) auxiliary Gaussian states dynamics induced by Eq.~\eqref{Lyapunov} along the isoergotropic surface $\mathcal{L}_{5}$, assuming $\hat{\rho}_{K}(0)=\hat{\rho}_{\bar{\mu}}(\mu_{K}(0),\xi, N_{K}(0))$ with $K=B, A$ (colored dots) as initial states, and $|\bar{\mu}|^2=5$, $N_A(0)=0$, $N_B(0)=0.8$ and $|\xi|=1$. Middle: Ergotropy and internal components dynamics for the (c) battery and (d) auxiliary system for the same parameters. Bottom: (e) Wigner function isoergotropic dynamics for the battery at instants $t=0,\pi/4,\pi/2$, assuming $\theta_B=0$, $\phi_B=\pi$ and $\omega=\eta=1$.\justifying}
    \label{fig10}
\end{figure}
\begin{figure}
    \center
    \includegraphics[width=\columnwidth]{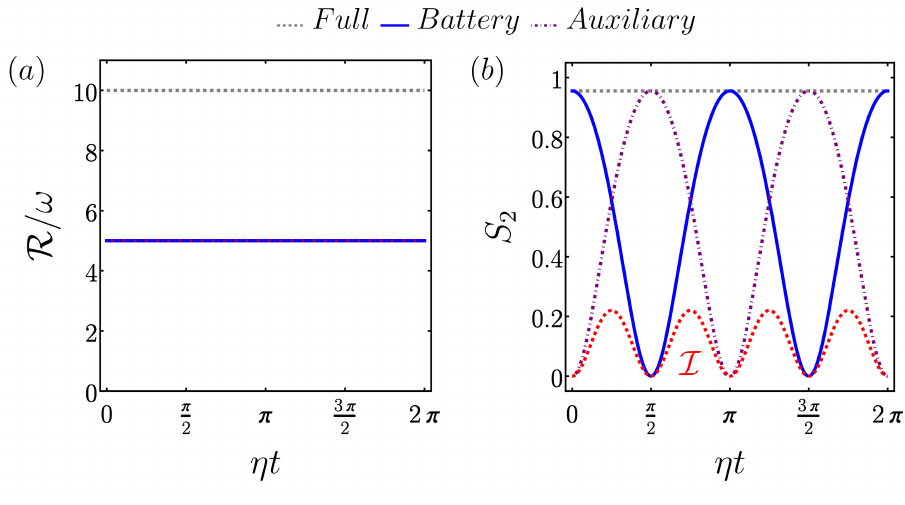}
    \caption{(a) Charge and (b) Rényi-2 entropy dynamics induced by Eq.~\eqref{LyapunovEquation}. The full (computed without the interaction) and local ergotropies are invariant during the time-evolution. The local Rényi-2 entropies and mutual information $\mathcal{I}$ oscillate during the internal conversion of ergotropy.\justifying}
    \label{fig11}
\end{figure}
%
%
\subsection{Dynamic ergotropy conversion}
Similarly to the TLS case, isoergotropic operations can be performed dynamically in a Gaussian quantum battery $(B)$ by coupling it to an auxiliary system $(A)$.
Assume two resonant bosonic modes with total Hamiltonian $\hat{H}=\hat{H}_{B}+\hat{H}_{A}+\hat{V}_{AB}$, where $\hat{H}_{B}=\omega(\hat{b}^{\dagger}\hat{b}+1/2)$, $\hat{H}_{A}=\omega\left(\hat{a}^{\dagger}\hat{a}+1/2\right)$ and a beam-splitter interaction
\begin{equation}\label{int2}
    \hat{V}_{AB}=i\eta\left(\hat{b}^{\dagger}\otimes\hat{a}-\hat{b}\otimes\hat{a}^{\dagger}\right),
\end{equation}
with $\eta$ being the coupling strength. Under these circumstances, the joint unitary dynamics is Gaussian and fully captured by the following Lyapunov equation
\begin{equation}\label{LyapunovEquation}
\begin{aligned}
\frac{d}{dt}\boldsymbol{D}(t)&=\boldsymbol{W}\boldsymbol{D}(t),\\\frac{d}{dt}\boldsymbol{\Xi}(t)&=\boldsymbol{W}\boldsymbol{\Xi}(t)+\boldsymbol{\Xi}(t)\boldsymbol{W}^{\dagger},
\end{aligned}
\end{equation}
for the first and second moments, with drift matrix $\boldsymbol{D}(t)=\left(\mu_{B}(t),\mu_{B}^{*}(t),\mu_{A}(t),\mu_{A}^{*}(t)\right)^{T}$, covariance matrix $\boldsymbol{\Xi}(t)$ at instant $t$, and
\begin{equation}
    \boldsymbol{W}=\left(\begin{array}{cccc}
-i\omega & 0 & \eta & 0\\
0 & i\omega & 0 & \eta\\
-\eta & 0 & -i\omega & 0\\
0 & -\eta & 0 & i\omega
\end{array}\right).
\end{equation}
Equation \eqref{LyapunovEquation} describes a beam-splitter rotation between $A$ and $B$ at rate $\eta$ while both modes freely precess at frequency $\omega$. Formally, the solution of Eq.~\eqref{LyapunovEquation} is given by
\begin{equation}
    \boldsymbol{D}(t)=\boldsymbol{\Lambda}(t)\boldsymbol{D}(0),\qquad\boldsymbol{\Xi}(t)=\boldsymbol{\Lambda}(t)\boldsymbol{\Xi}(0)\boldsymbol{\Lambda}^{\dagger}(t)
\end{equation}
where $\boldsymbol{\Lambda}(t)=e^{\boldsymbol{W}t}$.

In particular, assuming the initial isoergotropic Gaussian states $\hat{\rho}_{K}(0)=\hat{\rho}_{\bar{\mu}}(\mu_{K}(0),\xi,N_{K}(0))$ with $K=B,A$, such that
\begin{equation}
    \mu_{K}(0)=\sqrt{f_{\bar{\mu}}(\xi,N_{K}(0))}e^{i\theta_{K}},
\end{equation}
with $\theta_{A}=\theta_{B}+\pi/2$, and $\boldsymbol{\Xi}(0)=\boldsymbol{\Theta}_{B}(0)\oplus\boldsymbol{\Theta}_{A}(0)$ with
\begin{equation}
    \boldsymbol{\Theta}_{K}(0)=\left(N_{K}(0)+1/2\right)\boldsymbol{F}(\xi,\phi),
\end{equation}
one can show that
\begin{equation}\label{Lyapunov}
\begin{aligned}
\mu_{B}(t)&=e^{-it\omega}(\mu_{A}(0)\sin(\eta t)+\mu_{B}(0)\cos(\eta t)),\\\mu_{A}(t)&=e^{-it\omega}(\mu_{A}(0)\cos(\eta t)-\mu_{B}(0)\sin(\eta t)),
\end{aligned}
\end{equation}
and
\begin{equation}
    \boldsymbol{\Theta}_{K}(t)=\left(N_{K}(t)+1/2\right)\boldsymbol{F}(\xi,\phi-2t\omega),
\end{equation}
where
\begin{equation}
\begin{aligned}
N_{K}(t)&=\frac{1}{2}\left(N_{K}(0)-N_{K^{\prime}}(0)\right)\cos(2\eta t)\\&\quad+\frac{1}{2}\left(N_{A}(0)+N_{B}(0)\right)
\end{aligned}
\end{equation}
with $K^{\prime}\neq K$. Therefore, both battery and auxiliary systems remain within the same initial isoergotropic surface with fixed $\xi$, such that
\begin{equation}
    \hat{\rho}_{B(A)}(t)\equiv\hat{\rho}_{\bar{\mu}}(\mu_{B(A)}(t),\xi,N_{B(A)}(t))\in\mathcal{L}_{\bar{\mu}^{2}}.
\end{equation}
Figure~\ref{fig10} shows the dynamics of the (a) battery and (b) auxiliary system, assuming $|\bar{\mu}|^2=5$, $N_A(0)=0$, $N_B(0)=0.8$, and fixed $|\xi|=1$. The initial state of $B$ ($A$) is represented by the yellow (orange) dots. It is clear that, during the dynamics, both systems change within $\mathcal{L}_{5}$. While both local ergotropies are constant, the displacement and squeezing components oscillate with period $\eta T=\pi$. The former changes due to $\mu_{K}(t)$, while the latter is due to $N_{K}(t)$. 
Moreover, Fig.~\ref{fig10} (e) illustrates the Wigner function at different instants of time, for $\theta_B=0$, $\phi_B=\pi$, and $\omega=\eta=1$. For $t=0$, most of the battery's charge is stored in the squeezing component. In $t=\pi/4$, one observes $\mathcal{R}^{\textrm{d}}_{B}=\mathcal{R}^{\textrm{s}}_{B}$, until the ergotropy is transferred to the displacement component.
The auxiliary system follows the opposite behavior.
Notice that such a process has no internal cost of ergotropy, i.e., both the total and local charges are invariant, as shown by Fig.~\ref{fig11} (a). Thus, the only cost to realize such internal reorganization is due to the interaction term.
To evaluate how the correlations change, Fig.~\ref{fig11} (b) depicts how the Rényi-2 entropies and mutual information change during the dynamics.
%
%
%
%
%
%
%
%
%
%
%
%
%
%
%
%
%

%
%
\begin{figure}
    \center
    \includegraphics[width=0.8\columnwidth]{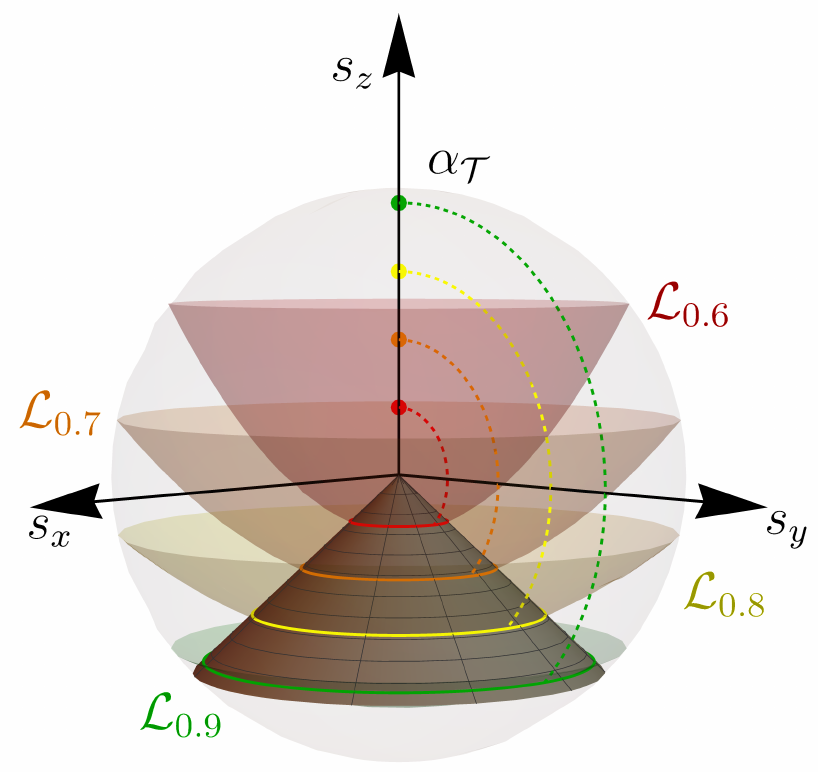}
    \caption{Isoergotropic surfaces and optimal-power charging. Starting from passive (uncharged) states, the cone marks all states reachable at maximal average power in the direct protocol; it corresponds to the fixed polar angle $\alpha_{\mathcal T}=\varepsilon\mathcal T\simeq 0.74\pi$. Colored dots and dashed arcs illustrate representative charging trajectories. The circular arcs are the intersections between the cone and the iso-$\mathcal R$ surfaces $\mathcal L_{\bar p}$; points on an intersection are optimal states $\hat\rho_{\bar p}(p,\theta)$ with $p=\tfrac12\!\left(1-\bar s\cos\alpha_{\mathcal T}\right)$ and $\bar s=P_{\bar p}\csc^{2}(\alpha_{\mathcal T}/2)$.\justifying}
    \label{figCharging}
\end{figure}
%
%
%
%
%
\section{Applications}\label{Application}
To highlight the potential significance of understanding the set of isoergotropic states, in the following, we will examine two relevant scenarios for quantum batteries: charging and discharging.  
\subsection{Charging}
\begin{figure*}
    \center
    \includegraphics[width=2\columnwidth]{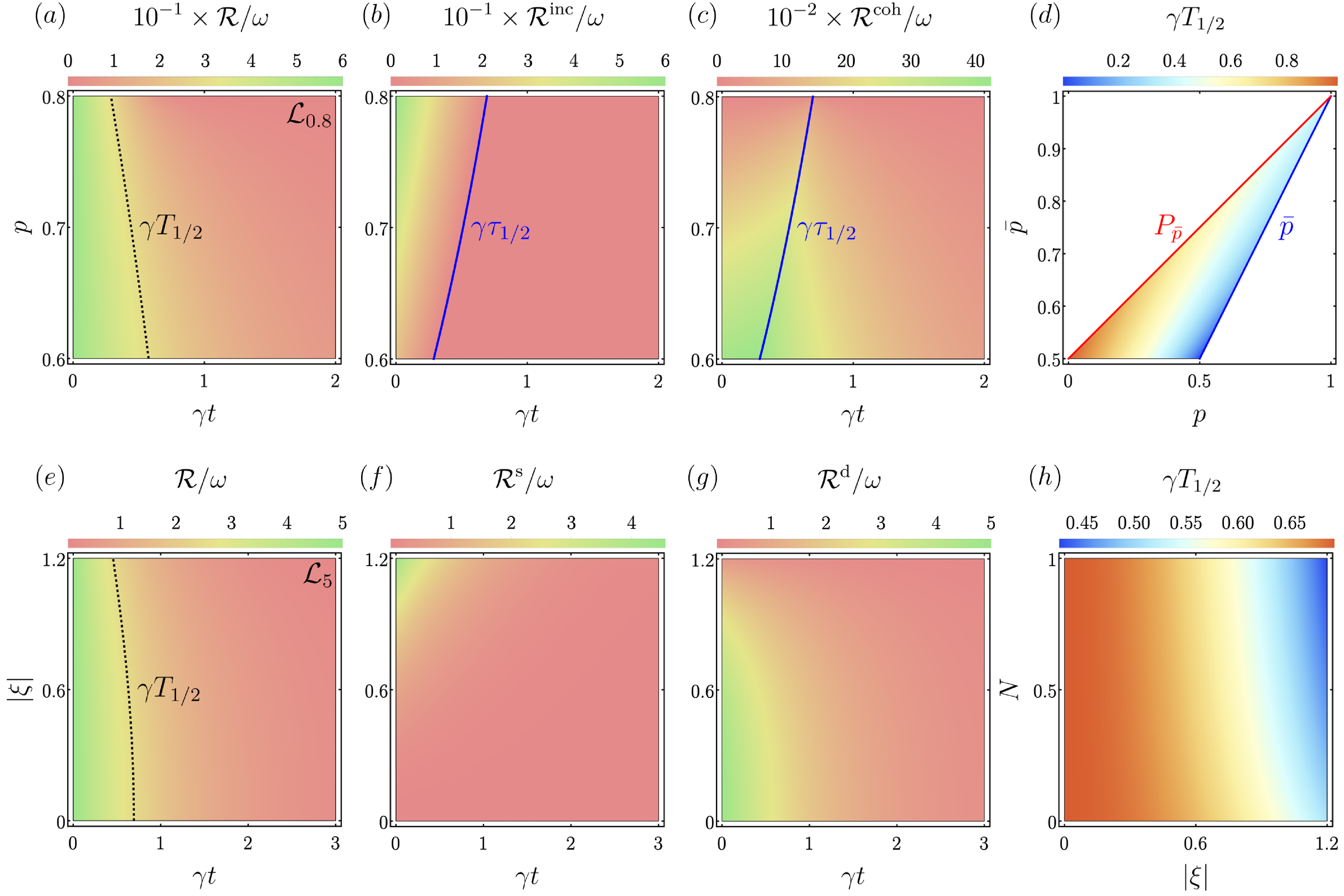}
    \caption{Ergotropy decay for (a)-(d) TLS (e)-(h) single-mode Gaussian-based open quantum batteries. (a) Ergotropy dynamics for the distinct isoergotropic states along $\mathcal{L}_{0.8}$. The black dashed line highlights the half-life of the battery. The (b) incoherent and (c) coherent components present distinct behavior. The blue continuous line indicates the instant with maximum coherent ergotropy, where $p_{\tau_{1/2}}=1/2$. As a result, states with coherence possess lower decay rates. The half-life behavior for all values of $\bar{p}$ is shown in (d). (e) Ergotropy dynamics for the distinct isoergotropic states along $\mathcal{L}_{5}$ and $N=0.5$. Both the (f) squeezing and (g) displacement components present a monotonic decay with time. Also, it is clear that states with higher values of $|\xi|$ decay faster. This behavior is observed for distinct values of $N$, as shown in (h). It was assumed (a)-(d) $\bar{p}=0.8$ $\bar{n}=0.2$, (e)-(h) $N=0.5$, $\bar{n}=0.3$ and $|\bar{\mu}|^2=5$.\justifying}
    \label{fig7}
\end{figure*}
Charging a quantum battery—i.e., increasing its ergotropy—amounts to transforming a passive state into an active one. In the simplest setting this can be done by coupling the storage to a semi-classical charger for a fixed duration $\mathcal T$. This direct charging protocol~\cite{PhysRevApplied.23.024010} is modeled by a time-dependent Hamiltonian $\hat H(t)=\hat H_B+\hat V(t)$, with $\hat H_B$ the battery Hamiltonian and $\hat V(t)$ the charging drive, switched on only for $0<t<\mathcal T$ ($\hat V(t)=0$ otherwise). The evolution is unitary, $\dot{\hat\rho}(t)=-i[\hat H(t),\hat\rho(t)]$, and supplies both energy $\langle \hat H_B\rangle$ and ergotropy $\mathcal R$ to the battery.
A natural performance goal is to maximize the average charging power under realistic constraints. We define
\begin{equation}\label{CharPow}
\langle P\rangle\coloneqq \frac{\langle W\rangle}{\mathcal T},\qquad
\langle W\rangle=\mathrm{Tr}\!\big[\hat H_B\big(\hat\rho(\mathcal T)-\hat\rho(0)\big)\big],
\end{equation}
and assume a bounded driving strength, $\|\hat H(t)\|\le \varepsilon$. For a single TLS with $\hat H_B=|e\rangle\langle e|$ and Bloch representation $\hat\rho(t) = \big(\hat{\mathds 1}+\boldsymbol s(t)\cdot\boldsymbol\sigma\big)/2$, $\boldsymbol s(t) = s\big(\sin\alpha_t\cos\varphi_t, \sin\alpha_t\sin\varphi_t, \cos\alpha_t\big)^{T}$, the optimal protocol with respect to the power under the norm bound is~\cite{binder2015quantacell}
\begin{equation}
\hat H(t)=-\frac{\varepsilon}{2}\sin\varphi_t\,\hat\sigma_x
+\frac{\varepsilon}{2}\cos\varphi_t\,\hat\sigma_y,
\end{equation}
which drives
\[
\varphi_t=\varphi_0,\qquad \alpha_t=\alpha_0+\varepsilon t.
\]
Starting from a passive (uncharged) state on the north pole,
$\alpha_0=0$ and $\boldsymbol s(0)=s_0(0,0,1)^{T}$ with $s_0\in[0,1]$,
the average power evaluates to $\langle P\rangle=s_0\big[1-\cos(\varepsilon\mathcal T)\big]/2\mathcal T$. Maximizing $\langle P\rangle$ over $\mathcal T$ yields
\begin{equation}
\cos(\varepsilon\mathcal T)+\varepsilon\mathcal T\,\sin(\varepsilon\mathcal T)=1,
\end{equation}
whose solution gives $\alpha_{\mathcal T}=\varepsilon\mathcal T\simeq 0.74\pi$. Thus the set of active states attainable with optimal power forms a cone on the Bloch sphere,
\[
\boldsymbol s_{\mathrm{Opt}}
=s\big(\sin\alpha_{\mathcal T}\cos\varphi,\ \sin\alpha_{\mathcal T}\sin\varphi,\ \cos\alpha_{\mathcal T}\big)^{T},
\]
with $s\in[0,1],\ \varphi\in[0,2\pi]$. Figure~\ref{figCharging} shows this optimal set together with several isoergotropic surfaces. Their intersections (solid colored curves) are parameterized by $\bar s=P_{\bar p}\csc^{2}(\alpha_{\mathcal T}/2)$ for different $\mathcal L_{\bar p}$. Consequently, for an initially uncharged battery the isoergotropic family reachable at maximum power is
\begin{equation}
\bar p=\frac12\left(1+\frac{s_0}{\csc^{2}(\alpha_{\mathcal T}/2)}\right),
\end{equation}
and the corresponding states $\hat\rho_{\bar p}(p,\theta)$ are given by
\begin{equation}
p=\frac12\Big(1-\bar s\,\cos\alpha_{\mathcal T}\Big),
\end{equation}
as indicated by the colored dots (initial states) and dashed lines (driven trajectories) in Fig.~\ref{figCharging}.
\subsection{Open QB and discharging}
External dissipation and decoherence degrade the performance of quantum devices in general~\cite{RevModPhys.95.045005,Galve2017,PhysRevA.105.012432,Molitor2024,aziz2025revival}. Mitigating ergotropy leakage is therefore essential for realistic quantum batteries~\cite{PhysRevResearch.2.013095,bv4w-jr6q}. The ability to reshuffle the internal composition of ergotropy, developed above, enables the preparation of states that are more robust against environmental noise.
To illustrate this, we couple the battery weakly to a thermal reservoir at inverse temperature $\beta$. The local dynamics of a TLS and of a single-mode Gaussian state (GS) are described by the GKLS master equation ($\hbar=k_B=1$)~\cite{breuer2002theory}:
\begin{equation}\label{ME}
\begin{aligned}
\textrm{TLS:}\,\dot{\hat{\rho}}&=-i[\hat{H},\hat{\rho}]+\gamma(1-\bar{n})\mathcal{D}_{\rho}[\hat{\sigma}_{+}]+\gamma\bar{n}\mathcal{D}_{\rho}[\hat{\sigma}_{-}],\\\textrm{GS:}\,\dot{\hat{\rho}}&=-i[\hat{H},\hat{\rho}]+\gamma(1+\bar{n})\mathcal{D}_{\rho}[\hat{a}]+\gamma\bar{n}\mathcal{D}_{\rho}[\hat{a}^{\dagger}],
\end{aligned}
\end{equation}
where $\gamma$ is the decay rate, $\bar n=(e^{\beta\omega}\pm1)^{-1}$ is the fermionic (bosonic) occupation number for the TLS (GS), and $\mathcal D_\rho[\hat o]=\hat o\hat\rho\hat o^\dagger-\{\hat o^\dagger \hat o,\hat\rho\}/2$. Equation~\eqref{ME} relaxes each system to the Gibbs state at temperature $1/\beta$, discharging the battery. Crucially, the coherent and incoherent (displacement and squeezing) parts of ergotropy decay differently, and the observed discharge reflects their interplay.

Regarding the TLS system at time $t$ the state has the form $\hat{\rho}(t)=\left(1-p_{t}\right)|g\rangle\langle g|+p_{t}|e\rangle\langle e|+\mathcal{C}_{t}\left(e^{i\frac{\theta}{2}}|g\rangle\langle e|+h.c.\right)/2$, with populations and coherence evolving independently as $p_{t}=\left(p_{0}-\bar{n}\right)e^{-\gamma t}+\bar{n}$ and $\mathcal{C}_{t}^{2}=\mathcal{C}_{0}^{2}e^{-\gamma t}$. Thus $p_\infty=\bar n$ and $\mathcal C_\infty=0$. The (in)coherent components of the ergotropy can be computed using Eqs.~\eqref{IncoherentErgoFormula} and~\eqref{CoherentErgoFormula}.
Interestingly, as observed by a previous work~\cite{bv4w-jr6q}, despite the monotonically decreasing behavior of the population and coherence, non-intuitive ergotropy dynamics can appear due to coherence, i.e., depending on the initial conditions, the coherent ergotropy becomes non-monotonic in time. More specifically, the maximum of $\mathcal{R}^{\textrm{coh}}$ is reached for $p=1/2$, thus if $p_{0}>1/2$, one observes a first increase of the coherent ergotropy until $p_{\tau_{1/2}}=1/2$ before asymptotically approaching zero, where $\tau_{1/2}\coloneqq\gamma^{-1}\ln\left[\left(p_{0}-\bar{n}\right)\left(1/2-\bar{n}\right)^{-1}\right]$ is the time required for the incoherent state become passive.
Figures~\ref{fig7}(a)–(c) show, for all initial states on $\mathcal L_{0.8}$, that the total and incoherent charges decay monotonically, while the coherent part can exhibit a transient increase. Panel~\ref{fig7}(a) marks the half-life $T_{1/2}$ defined by $\mathcal R(T_{1/2})=\mathcal R(0)/2$. States with larger initial coherence have longer $T_{1/2}$ despite sharing the same initial charge. Figure~\ref{fig7}(d) depicting the half-life time in terms of $\bar{p}$ and $P_{\bar{p}}\leq p \leq \bar{p}$ confirms this trend across different iso-$\mathcal R$ surfaces: within a given $\mathcal L_{\bar p}$, coherent (near-pure) states on the Bloch-sphere surface are the most resilient to environment-induced dissipation.

As for the GS states, since the dissipators in Eq.~\eqref{ME} are linear in $\hat a$ and $\hat a^\dagger$, Gaussianity is preserved. The ergotropy at time $t$ is given by Eq.~\eqref{ErgoGaussian} with time-dependent parameters $(\mu_t,\xi_t,N_t)$ evolving as~\cite{medina2024anomalous}
\begin{equation}
\begin{aligned}
|\mu_{t}|^{2}&=|\mu_{0}|^{2}e^{-\gamma t},\\|\xi_{t}|&=\frac{1}{2}\textrm{cosh}^{-1}\frac{\left[\Delta_{t}+\left(2N+1\right)e^{-\gamma t}\sinh^{2}\left(|\xi_{0}|\right)\right]}{N_{t}+1/2},\\N_{t}&=\sqrt{\Delta_{t}^{2}+\frac{\left(2N+1\right)\left(2\bar{n}+1\right)}{e^{\gamma t}(1-e^{-\gamma t})^{-1}}\sinh^{2}\left(|\xi_{0}|\right)}-\frac{1}{2},
\end{aligned}
\end{equation}
with $\Delta_{t}=2\left(N-\bar{n}\right)e^{-\gamma t}+\left(\bar{n}+1/2\right)$.
Figures~\ref{fig7}(e)–(g) display the decay of charge for different initial points on $\mathcal L_{5}$ with $N=0.5$: both the squeezing and displacement contributions decrease monotonically, but the half-life is shorter for larger initial $|\xi|$. This “hotter empties faster” behavior is the ergotropic analogue of the Mpemba effect~\cite{medina2024anomalous}. Panel~\ref{fig7}(h) shows $T_{1/2}$ versus $|\xi|$ for several $N$, confirming the trend across $\mathcal L_{5}$.

In summary, consistent with Refs.~\cite{bv4w-jr6q,medina2024anomalous}, tailoring the internal composition of ergotropy—either at preparation (charging) or via iso-$\mathcal R$ operations—yields states that retain charge longer under open-system dynamics, providing a practical advantage for quantum batteries in realistic noisy environments.
%
%
%

\section{Discussion}\label{DiscussionSection}
A central challenge in the study of quantum batteries (QBs) lies not only in maximizing their stored energy but also in understanding how this charge is internally structured and can be effectively manipulated. While ergotropy $\mathcal{R}$ provides a fundamental measure of the extractable work from a quantum state, recent insights show that this quantity can be distributed among distinct internal components, classical and quantum degrees of freedom. This internal composition plays a crucial role in determining how the stored charge can be accessed, transformed, or preserved during realistic operations, including efficient charging and work extraction.

In this broader context, we introduced and investigated the set $\mathcal{L}$ of isoergotropic states, defined as states possessing the same total ergotropy but different internal configurations of ergotropy. Although such states are equivalent in terms of their stored charge, they may exhibit markedly distinct physical properties, including internal energy, purity, and entropy. We further explored the class $\mathcal{K}$ of quantum operations—comprising completely positive trace-preserving (CPTP) maps and selective measurements that act within $\mathcal{L}$, preserving total ergotropy while redistributing its internal components. 
The thermodynamics involved during such processes is also verified, including heat transfer and the possibility of isentropic and internal energy-conserving transformations.
As relevant examples, these concepts are illustrated and applied with both discrete and continuous-variable quantum batteries, respectively represented by two-level systems and Gaussian states. While in the former, the charge is stored within the coherent and incoherent components, in the latter, it is independently distributed along the displacement and squeezing ones.

We also demonstrated that dynamical conversion between the internal components of ergotropy can be achieved by coupling the quantum battery to an auxiliary system. In particular, Eqs.~\eqref{int1} and~\eqref{int2}, describing the interactions for the TLS and Gaussian cases, share the structure of a beam-splitter-type coupling. Such interactions enable coherent energy exchange while conserving the total number of excitations in the composite system.
These couplings allow the autonomous reconfiguration of ergotropy and the simultaneous realization of isoergotropic transformations within the target systems. Based on the results reported in~\cite{PhysRevLett.133.180401}, we anticipate that these processes could be exploited to selectively assess and extract specific components of the stored charge.
Beam-splitter-type interactions are standard tools in quantum optics and are widely used across various quantum technologies~\cite{math10244794,PhysRevA.110.023722,PhysRevA.111.053702}. Moreover, such couplings can be readily engineered in both optical and superconducting platforms~\cite{PhysRevX.8.021073,PRXQuantum.4.020355,10.1063/5.0272687}. Therefore, isoergotropic operations can be realistically implemented in quantum batteries using current state-of-the-art technologies.

To this end, we also explored the practical implications of iso-$\mathcal{R}$ states and operations in relevant scenarios. Considering a direct charging protocol for a TLS, we characterized the set of states that achieve optimal charging power and identify the corresponding optimal isoergotropic states within each $\mathcal{L}_{\bar{p}}$, as illustrated by the intersections in Fig.~\ref{figCharging}. In this way, one can either tune the initial discharged state $\hat{\rho}_0$ to reach a desired iso-$\mathcal{R}$ surface or determine the optimal charged state attainable from a given $\hat{\rho}_0$. Similar analyses can be performed for other optimization criteria and charging strategies.
Furthermore, we investigated the robustness of different isoergotropic states under environmental noise. By accounting for the distinct dynamical behavior of the ergotropic components, one can mitigate dissipative charge loss in open quantum batteries—either by charging into suitable isoergotropic states or by dynamically reshuffling the ergotropy between components. As shown in Fig.~\ref{fig7}(a), isoergotropic states with coherence exhibit longer half-life times $T_{1/2}$, thus decaying more slowly. A similar advantage is observed in Fig.~\ref{fig7}(b), where maximizing the ergotropy stored in the displacement component enhances robustness against dissipation.

The formalism and analyses developed here can be systematically extended to arbitrary quantum battery architectures. Future work should aim to identify additional applications and explicitly incorporate iso-$\mathcal{R}$ operations into practical protocols. For example, optimal charge stabilization strategies can be found through quantum control techniques for different relevant dynamical scenarios~\cite{kurizki2022thermodynamics}, including non-Markovian environments. 
Additionally, they can be utilized in the selective charging and extraction of specific ergotropic components. Such operations could be harnessed to design novel thermodynamic cycles whose strokes internally redistribute charge while maintaining a fixed energy budget. Also, one can study the potential use of non-isoergotropic local operations along with global isoergotropic processes in multi-cell quantum batteries. 
Finally, by taking isoergotropic transformations as free operations, one can build a resource theory of ergotropy in direct analogy with standard quantum resource theories (see the review~\cite{RevModPhys.91.025001}). Framed this way, ergotropy can be integrated with existing resource theories of coherence~\cite{RevModPhys.89.041003} and Gaussianity
\cite{PhysRevA.98.022335}, enabling analyses beyond the usual settings. Notably, recent work has designed nonlinear coherent heat machines that convert passive Gaussian states into non-Gaussian active ones~\cite{doi:10.1126/sciadv.adf1070}. This indicates that nonlinearity cannot be treated as a free resource in any continuous-variable resource theory of ergotropy. Investigating the interplay among these resources—and identifying appropriate monotones and conversion rules—would be valuable, but lies beyond the scope of the present work.

%

\section{Acknowledgements}
A.H.A.M. acknowledges support from National Science Centre, Poland Grant OPUS-21 (No. 2021/41/B/ST2/03207). B.A. and P.H. acknowledge support from IRA Programme (project no. FENG.02.01-IP.05-0006/23) financed by the FENG program 2021-2027, Priority FENG.02, Measure FENG.02.01., with the support of the FNP. P.R.D. acknowledges support from the NCN Poland, ChistEra-2023/05/Y/ST2/00005 under the project Modern Device Independent Cryptography (MoDIC).
%
%
%
%
%
%
%
%
%
%
%
%
%
%
%
%
%
%
%
%
%
%
%
%
%
%
%
%
%
%
%
%
%
%
%
%
%
%
%
%
%
%
%
%
%
%
%
%
%
%
%
%
%
%
%

%
%

%
%

\bibliography{References}
\end{document}